\documentclass[12pt,preprint]{aastex}
\usepackage{tipa}
\usepackage{txfonts}
\usepackage{amssymb}

\begin{document}

\title{Comparison between RHD simulation of supercritical accretion flows and steady model with outflows}

\author{CHENG-LIANG JIAO,\altaffilmark{1,2}
        SHIN MINESHIGE,\altaffilmark{3}
        SHUN TAKEUCHI,\altaffilmark{3}
        and
        KEN OHSUGA,\altaffilmark{4, 5}
        }
\altaffiltext{1}{Yunnan Observatories, Chinese Academy of Sciences, Kunming 650216, Yunnan, China}
\altaffiltext{2}{Key Laboratory for the Structure and Evolution of Celestial Objects, Chinese Academy of Sciences, Kunming 650216, Yunnan, China}
\altaffiltext{3}{Department of Astronomy, Graduate School of Science, Kyoto University, Sakyo-ku, Kyoto 606-8502, Japan}
\altaffiltext{4}{National Astronomical Observatory of Japan, Osawa, Mitaka, Tokyo 181-8588, Japan}
\altaffiltext{5}{School of Physical Sciences, Graduate University of Advanced Study (SOKENDAI), Shonan Village, Hayama, Kanagawa 240-0193, Japan}
\email{E-mail : jiaocl@ynao.ac.cn}

\newcommand{\bm}[1]{\mbox{\boldmath $#1$}}

\begin{abstract}
We apply our two-dimensional (2D), radially self-similar steady-state accretion flow model to the analysis of hydrodynamic simulation results of supercritical accretion flows.  Self-similarity is checked and the input parameters for the model calculation, such as advective factor and heat capacity ratio, are obtained from time-averaged simulation data. Solutions of the model are then calculated and compared with the simulation results. We find that in the converged region of the simulation, excluding the part too close to the black hole, the radial distribution of azimuthal velocity $v_\phi$, density $\rho$ and pressure $p$ basically follows the self-similar assumptions, i.e. they are roughly proportional to $r^{-0.5}$, $r^{-n}$, and $r^{-(n+1)}$, respectively, where $n\sim0.85$ for the mass injection rate of $1000L_\mathrm{E}/c^2$, and $n\sim0.74$ for $3000L_\mathrm{E}/c^2$. The distribution of $v_r$ and $v_\theta$ agrees less with self-similarity, possibly due to convective motions in the $r\theta$ plane. The distribution of velocity, density and pressure in $\theta$ direction obtained by the steady model agrees well with the simulation results  {within the calculation boundary of the steady model}. Outward mass flux in the simulations is overall directed toward polar angle of 0.8382 rad ($\sim 48.0^\circ$) for $1000L_\mathrm{E}/c^2$, and 0.7852 rad ($\sim 43.4^\circ$) for $3000L_\mathrm{E}/c^2$, and $\sim$94\% of the mass inflow are driven away as outflow, while outward momentum and energy fluxes are focused around the polar axis. Part of these fluxes lie in the region that are not calculated by the steady model, and special attention should be paid when the model is applied.

\end{abstract}


\section{Introduction}
Recent development in observations has shown growing evidence that a large portion of mass is blown away from accretion flow onto black holes in the form of outflows. For example, the accretion rate onto the Galactic Center is estimated to be about $10^{-6}$ $M_\sun\ \mathrm{yr}^{-1}$ (Baganoff et al. 2003) at the outer boundary, while the detected high linear polarization at radio waveband limits the mass inflow rate near the event horizon to be $\lesssim10^{-8}$ $M_\sun\ \mathrm{yr}^{-1}$ (Baganoff et al. 2003) or $10^{-7}-10^{-9}$ $M_\sun\ \mathrm{yr}^{-1}$ (Marrone et al. 2007), which implies that most of the accreting material cannot reach the black hole. Outflows are also observed through blue-shifted absorption lines of galactic sources (e.g., Miller et al. 2004; Kotani et al. 2006; Kubota et al. 2007; Neilsen et al. 2012; Ponti et al. 2012) and active galactic nuclei (AGNs; Terashima \& Wilson 2001; Pounds et al. 2003; Reeves et al. 2003; Ganguly \& Brotherton 2008; Pounds \& Reeves 2009).

Theoretically, it is believed that outflows are likely to be generated when the mass supply rate is much less than ('underfed') or much greater than ('overfed') the Eddington accretion rate. The 'underfed' case can be described by an optically thin advection-dominated accretion flow (ADAF; Narayan \& Yi 1994, 1995a, b; Abramowicz et al. 1995; for reviews, see Narayan et al. 1998; Kato et al 2008; Narayan \& McClintock 2008). In an ADAF, energy released via viscous dissipation cannot be radiated away efficiently, which causes the gas to be heated to very high temperature and become unbound, and strong outflows are likely to be generated (e.g. Narayan \& Yi 1994; Quataert \& Narayan 1999). The 'overfed' case can be described by a 'slim disk' (Abramowicz et al. 1988; Beloborodov 1998; Chen \& Wang 2004; S{\c a}dowski 2009; Takeuchi et al. 2009; etc.). While the original slim disk model does not account for outflows due to treatment limit, it has been discovered that there is an upper limit of mass supply rate for slim disks beyond which outflows become inevitable (Gu \& Lu 2007; Jiao et al. 2009; Jiao \& Lu 2009; Gu 2012, 2015). In fact, even for accretion rates below this limit, outflows are likely to be generated due to strong radiation pressure (e.g. Ohsuga et al. 2009). In both cases, the existence of outflows has been validated by numerical simulations (e.g., Stone et al. 1999; Igumenshchev \& Abramowicz 2000; Stone \& Pringle 2001; Hawley et al. 2001; Hawley \& Balbus 2002; McKinney \& Gammie 2002; Igumenshchev et al. 2003; De Villiers et al. 2003; Okuda et al. 2005; Ohsuga et al. 2005, 2009; Ohsuga \& Mineshige 2007, 2011; Yuan \& Bu 2010; McKinney et al. 2012; Narayan et al. 2012; Yuan et al. 2012a, 2012b; Takeuchi et al. 2013; etc.).

However, it is very difficult to apply numerical simulation results directly to observations, e.g. fitting the spectra of accretion-powered astrophysical
systems, due to the heavy computation they require. Simple analytic steady models are still the only accessible way of making direct link between theory and observations for astronomers. Nevertheless, steady models which contain outflows are quite limited, among which are contributions from Narayan \& Yi (1995a), Xu \& Chen (1997), Blandford \& Begelman (1999, 2004; hereafter BB99, BB04), Xue \& Wang (2005), Gu et al.(2009), S{\c a}dowski et al.(2011), Jiao \& Wu (2011; hereafter JW11), Begelman (2012), etc. (see JW11 for a detailed discussion of the contributions and caveats of some of these works). Although these models are established self-consistently, it is not clear whether the assumptions and results of them agree with the simulation results, which in some sense represent the structure of real astrophysical accretion flows.

In this paper, we contribute to this topic by applying our 2D self-similar accretion flow model (JW11) to the analysis of simulation results. As the first step, we select two samples of 2D radiation-hydrodynamic (RHD) simulation of supercritical accretion flows (Ohsuga et al. 2005, hereafter OMNM05), in which the viscous stress is expressed with the $\alpha p$ prescription, similar to our steady model. In Section 2, we describe the simulation model and steady model used in this paper. In Section 3, we check the self-similar assumptions by analysing the time-averaged simulation data in quasi-steady state. In Section 4, we compare the steady model solutions with the simulation results. Discussions are presented in Section 5 and we conclude with a summary in Section 6.
\section{Models}
In this section we briefly describe the simulation model and the steady model used in this paper. More details of the equations and numerical methods can be found in respective papers (OMNM05 \& JW11).
\subsection{The simulation model}
The simulation is performed with the two-dimensional RHD code developed by Ohsuga et al.(2005). The calculation is carried out in spherical coordinates ($r,\theta,\phi$), with a non-rotating black hole at the origin. The accretion flow is assumed to have axisymmetry (i.e. $\partial/\partial\phi=0$) and reflection symmetry relative to the equatorial plane ($\theta=\pi/2$). Pseudo-Newtonian potential is adopted, which is given by $\psi=-GM/(r-r_g)$ (Paczy\'{n}ski \& Wiita 1980), in which $M$ is the black hole mass and $r_g=2GM/c^2$ is the Schwarzschild radius. Radiative transfer is described by the flux-limited diffusion (FLD) approximation (Levermore \& Pomraning 1981). It is assumed that the $r\phi$-component of
the viscous tensor, $t_{r\phi}$, is dominant.
The dynamical viscosity coefficient is given by
\begin{equation}\label{}
    \eta=\alpha\frac{p_\mathrm{gas}+\lambda E_\mathrm{rad}}{\Omega_\mathrm{K}},
\end{equation}
where $\alpha$ is the viscosity parameter, $\Omega_\mathrm{K}$ is the Keplerian angular speed, $\lambda$ is the flux limiter and $E_\mathrm{rad}$ is the radiation energy density. It is basically the same as the $\alpha$ prescription of the viscosity (Shakura \& Sunyaev 1973), because $\lambda$ is almost 1/3 in the optically thick region, and the results show that for supercritical accretion flows the majority of the flow in the converged region is optically thick.

The computational domain is set to be spherical shells of $3r_g \leq r \leq 500r_g$ and $0 \leq \theta \leq 0.5\pi$, and is divided into $96\times96$ grid cells. The calculation is started with a hot, rarefied and optically thin atmosphere with no initial cold dense disk, and mass is injected continuously into the computational domain through the outer boundary ($r=500r_g, 0.45\pi \leq \theta \leq 0.5\pi$) at a steady rate $\dot{M}_\mathrm{input}$. The specific angular momentum of the injected mass is set to be the same as the Keplerian angular momentum at $r=100r_g$. In this paper, two sets of simulation results are analyzed, both of which are taken from OMNM05 and have reached quasi-steady state for $3r_g \leq r \leq 100r_g$. For both simulation runs, the parameters are set as $M=10M_\sun$, $\alpha=0.1$, $\gamma=5/3$, $\mu=0.5$ and $Z=Z_\sun$, where $\gamma$, $\mu$ and $Z$ are the heat capacity ratio, mean molecular weight and metallicity, respectively. The mass injection rate is set to be $1000L_\mathrm{E}/c^2$ and $3000L_\mathrm{E}/c^2$ respectively, where $L_\mathrm{E}$ is the Eddington luminosity.

\subsection{The steady model}
The steady accretion disk model with outflows is basically the same as presented in JW11. The calculation is carried out in spherical coordinates ($r,\theta,\phi$) with the gravitational center at the origin, and the flow is assumed to be steady($\partial /\partial t=0$) and axisymmetric($\partial / \partial \phi=0$). The Newtonian gravitational potential, $\Phi=-GM/r$ is adopted, and the viscosity is described by the $\alpha$ prescription $t_{r\phi}=-\alpha p$, where $p=p_\mathrm{gas}+p_\mathrm{rad}$ is the total pressure (magnetic pressure is not considered). It is assumed that the $r\phi$-component of
the viscous tensor, $t_{r\phi}$, is dominant. The energy equation is described by the advective factor, $f\equiv Q_\mathrm{adv}/Q_\mathrm{vis}$, so that a fraction $f$ of the
dissipated energy is advected as stored entropy and a fraction $(1-f)$ is lost due to radiation. Note that Newtonian potential, rather than the pseudo-Newtonian potential, is adopted here, so later in the comparison part, we will focus on the region not too close to the central black hole ($r \geq 10 r_g$), where relativistic effects can be neglected.

Self-similar assumptions are adopted in the radial direction (Narayan \& Yi 1995a; Xue \& Wang 2005; etc.):
\begin{eqnarray}
  \rho &=& \rho(\theta)r^{-n}, \\
  v_r &=& v_r(\theta)\sqrt{\frac{GM}{r}}, \\
  v_\theta &=& v_\theta(\theta)\sqrt{\frac{GM}{r}}, \\
  v_\phi &=& v_\phi(\theta)\sqrt{\frac{GM}{r}}, \\
  p &=& p(\theta)GMr^{-n-1},
\end{eqnarray}
and the hydrodynamic equations can be reduced to a set of ordinary differential equations (ODEs) about the variable $\theta$. This set of ODEs can be numerically solved with the symmetric boundary conditions relative to the equatorial plane if $\rho(\pi/2)$ is set to be 1, normalized by a scale factor if the effective accretion rate at a certain radius is set (Narayan \& Yi 1995a; Xue \& Wang 2005; etc.). Four input parameters
($\alpha$,$f$,$\gamma_\mathrm{equ}$,$n$) are required to calculate a solution, in which $\gamma_\mathrm{equ}$ is defined as
\begin{equation}\label{}
    \gamma_\mathrm{equ}\equiv \frac{\gamma-1}{\beta+3(1-\beta)(\gamma-1)}+1,
\end{equation}
where $\beta \equiv p_\mathrm{gas}/p$ is the gas pressure ratio.  {$\gamma_\mathrm{equ}$ is defined to incorporate the influences of both gas and radiation pressure into the model. For pure ionized hydrogen ($\gamma=5/3$), in the case of extreme gas pressure domination ($\beta \rightarrow 1$), $\gamma_\mathrm{equ}=5/3$, while in the case of extreme radiation pressure domination ($\beta \rightarrow 0$), $\gamma_\mathrm{equ}=4/3$.}

 {The self-similar steady model here cannot describe the accretion flow structure in the whole space. Theoretically, this is because that, with the self-similar assumptions, the inflow accretion rate and outflow accretion rate scale with radius in the same way (both are proportional to $r^{1.5-n}$). In the steady state, the total accretion rate should remain constant, which does not scale with radius. If the whole space can be described by the self-similar model, then there must be n=1.5 and both inflow and outflow accretion rate have to be constant. This means that there is no outflow (as outflow equals the difference between inflow accretion rates at different radii).  In fact the n=1.5 case has been solved by Narayan \& Yi (1995a), in which all the streamlines are straight lines pointing at the central accretor and no outflow exists. So the self-similar steady model which contains outflow must have some boundary beyond which it cannot describe. Numerically, the calculation starts from the equatorial plane ($\theta=\pi/2$) and moves towards the polar axis ($\theta=0$). Both $p$ and $\rho$ decreases as $\theta$ decreases, and at some polar angle they will get very close to 0 simultaneously. If we continue the calculation beyond this angle, we will encounter numerical errors, so we take this polar angle $\theta$ as the upper boundary of the steady model.}

It is worth mentioning that out of the four input parameters ($\alpha$,$f$,$\gamma_\mathrm{equ}$,$n$), $f$ does not appear in any differential terms in the equations, and thus can actually be variant in the $\theta$ direction. The other three parameters are included in differential terms  {(see JW11 for detailed equations)}, but can also be set as functions of $\theta$, as long as the velocities, density and pressure do not change abruptly in space in the solution. To compare the steady model results with the simulation results, we obtain the input parameters ($\alpha$,$f$,$\gamma_\mathrm{equ}$,$n$) from the simulation data, which will be described in detail later.

Here we adopt the self-similar assumptions presented in Narayan \& Yi (1995a) and Xue \& Wang (2005). It should be noted that in Narayan \& Yi (1995a), the self-similar form of $v_\phi$ is actually
\begin{equation}\label{ny95}
    v_\phi=v_\phi(\theta)\sin{\theta}\sqrt{\frac{GM}{r}},
\end{equation}
although in their paper it is stated otherwise (possibly a typo; also see Tanaka \& Menou 2006). If we use Eq. (\ref{ny95}) to substitute Eq. (5), and leave other self-similar forms unchanged, we can write another version of the code. We run both versions of the code and find that with the same input parameters ($\alpha$,$f$,$\gamma_\mathrm{equ}$,$n$), the results are actually the same. It is not surprising because Eq.(5) is in fact equivalent to Eq.(\ref{ny95}) if we include the $\sin{\theta}$ term into the $v_\phi(\theta)$ term.
\section{Checking the self-similar assumptions}
In this section we compare the self-similar assumptions used in the steady model with the simulation data.
From Eqs.(2)-(6), we can get the following relations (for brevity in this paper we use 'lg' to represent the logarithm with base 10):
\begin{equation}\label{}
    \lg{v_r}=-\frac{1}{2}\lg{r}+c_1(\theta),
\end{equation}
\begin{equation}\label{}
    \lg{v_\theta}=-\frac{1}{2}\lg{r}+c_2(\theta),
\end{equation}
\begin{equation}\label{}
    \lg{v_\phi}=-\frac{1}{2}\lg{r}+c_3(\theta),
\end{equation}
\begin{equation}\label{}
    \lg{\rho}=-n\lg{r}+c_4(\theta),
\end{equation}
\begin{equation}\label{}
    \lg{p}=-(n+1)\lg{r}+c_5(\theta).
\end{equation}
Here $c_1, c_2, c_3, c_4$ and $c_5$ are dependant on the polar angle $\theta$, while they should remain constant for a fixed $\theta$ under the self-similar assumptions for a certain steady model solution. In JW11, we assume that $n$ is constant, which means that for these relations the slopes should not change at different $\theta$. The simulation data, on the other hand, do not adopt these assumptions, and do not necessarily follow these relations. Here we fit the simulation data with linear models to check whether the self-similar assumptions in radial direction are in good agreement with the simulation results. The data are averaged over $t=185-255$ s of the simulation for $\dot{M}_\mathrm{input}=1000L_\mathrm{E}/c^2$ and $t=182-252$s for $\dot{M}_\mathrm{input}=3000L_\mathrm{E}/c^2$, during which the simulations are in quasi-steady state, respectively. As we are interested in the converged part of the simulation data except for the part very close to the black hole, we focus on the region from 10$r_g$ to 100$r_g$ in the simulation data. This correspond to the 23rd grid point to the 67th grid point. All the 96 rows divided in $\theta$ direction will be taken into consideration.

The detailed fitting results and some discussion are presented in the following subsections. To evaluate the curve fitting results, we make visual examination of the fitted curves, as well as focus on two quantities: the fitted slopes and $R^2$ of fits. The fitted slopes are constant for velocities and directly connected with the input parameter $n$ for density and pressure, and should remain constant for different $\theta$ for a certain fit, as discussed above. $R^2$ is a goodness-of-fit statistic which is defined as
\begin{equation}\label{}
  R^2=1-\frac{\sum\limits_{i}(y_i-\hat{y}_i)^2}{\sum\limits_{i}(y_i-\bar{y})^2},
\end{equation}
where $y_i$ is the value of a physical quantity from the simulation data, $\hat{y}_i$ is the fitted value from the linear regression, and $\bar{y}$ is the mean of $y_i$.
$R^2$ measures how well the regression line approximates the simulation data points(the closer to 1, the better; an $R^2$ of 1 indicates that the regression line perfectly fits the data.). Table 1 summarizes all the fitting results.

\begin{table}
\begin{center}
\caption{Linear Fitting Results of Simulation Data (vs $\lg{r}$)}
\begin{tabular}{lcccc}
\tableline\tableline
 Variables & \multicolumn{4}{c}{$\dot{m}=1000$} \\
\cline{2-5} \\
 & $R^2$ & $\bar{R^2}$ & Slope & Mean Slope\\
\tableline
$\lg{v_r}$(inflow) & 0.00$\ \sim\ $0.64 & 0.31 & -0.94$\ \sim\ $0.44 & -0.27\\
$\lg{v_r}$(outflow) & 0.15$\ \sim\ $0.80 & 0.36 & 0.07$\ \sim\ $0.88 & 0.36\\
$\lg{v_\theta}$ & 0.21$\ \sim\ $0.95 & 0.81 & -1.34$\ \sim\ $0.36 & -0.84\\
$\lg{v_\phi}$ & 0.98$\ \sim\ $0.99 & 0.99 & -0.75$\ \sim\ $-0.56 & -0.65\\
$\lg{\rho}$ & 0.87$\ \sim\ $0.99 & 0.94 & -1.43$\ \sim\ $-0.76 & -0.85\\
$\lg{p}$ & 0.98$\ \sim\ $0.99 & 0.99 & -2.10$\ \sim\ $-1.72 & -2.03\\

\tableline
Variables & \multicolumn{4}{c}{$\dot{m}=3000$} \\
\cline{2-5} \\
 & $R^2$ & $\bar{R^2}$ & Slope & Mean Slope\\
\tableline
$\lg{v_r}$(inflow) & 0.00$\ \sim\ $0.91 & 0.59 & -1.20$\ \sim\ $0.65 & -0.44\\
$\lg{v_r}$(outflow) & 0.10$\ \sim\ $0.82 & 0.31 & 0.11$\ \sim\ $0.86 & 0.38\\
$\lg{v_\theta}$ & 0.33$\ \sim\ $0.98 & 0.82 & -1.09$\ \sim\ $0.24 & -0.74\\
$\lg{v_\phi}$ & 0.97$\ \sim\ $0.99 & 0.99 & -0.73$\ \sim\ $-0.53 & -0.64\\
$\lg{\rho}$ & 0.85$\ \sim\ $0.99 & 0.94 & -1.31$\ \sim\ $-0.55 & -0.74\\
$\lg{p}$ & 0.98$\ \sim\ $0.99 & 0.99 & -2.02$\ \sim\ $-1.59 & -1.92\\

\tableline
\tableline
\end{tabular}
\end{center}
\end{table}

\subsection{Fitting velocities}
The fitting results of $v_r, v_\theta$ and $v_\phi$ are presented in this subsection. Linear fits have been performed at all inclination angles on the simulation grid, except for $v_\theta$ on the equatorial plane where $v_\theta=0$ due to reflection symmetry of the accretion flow. In this paper we define $\dot{m} \equiv \dot{M}_\mathrm{input}/(L_\mathrm{E}/c^2)$ for brevity. All the physical quantities are in cgs units.

 {Figure 1 shows the linear fits of $\lg{|v_r|}$, $\lg{|v_\theta|}$ and $\lg{v_\phi}$ at polar angle $\theta=1.2732$ rad in the inflow region, and Figure 2 shows the fits at $\theta=\pi/4$ in the outflow region.} We fit the absolute values of $v_r$ and $v_\theta$ here, because in the simulation data there exist circulation patterns in the velocity fields in the $r\theta$ plane, which is due to convection (see OMNM05).  {Figure 3 gives a general view of the fitting results of $\lg{|v_r|}$, $\lg{|v_\theta|}$ and $\lg{v_\phi}$ at different polar angles.} The upper panel corresponds to the $R^2$ values, which are better for values closer to 1. The lower panel corresponds to the fitted slopes, which are better for values deviating less from each other on the same curve.  {For the fits of $\lg{|v_r|}$, we ignore $\theta$ values for which there exists $v_r=0$, and the left branch corresponds to the outflow region, while the right branch corresponds to the inflow region.}

\begin{figure}[htbp]
\plotone{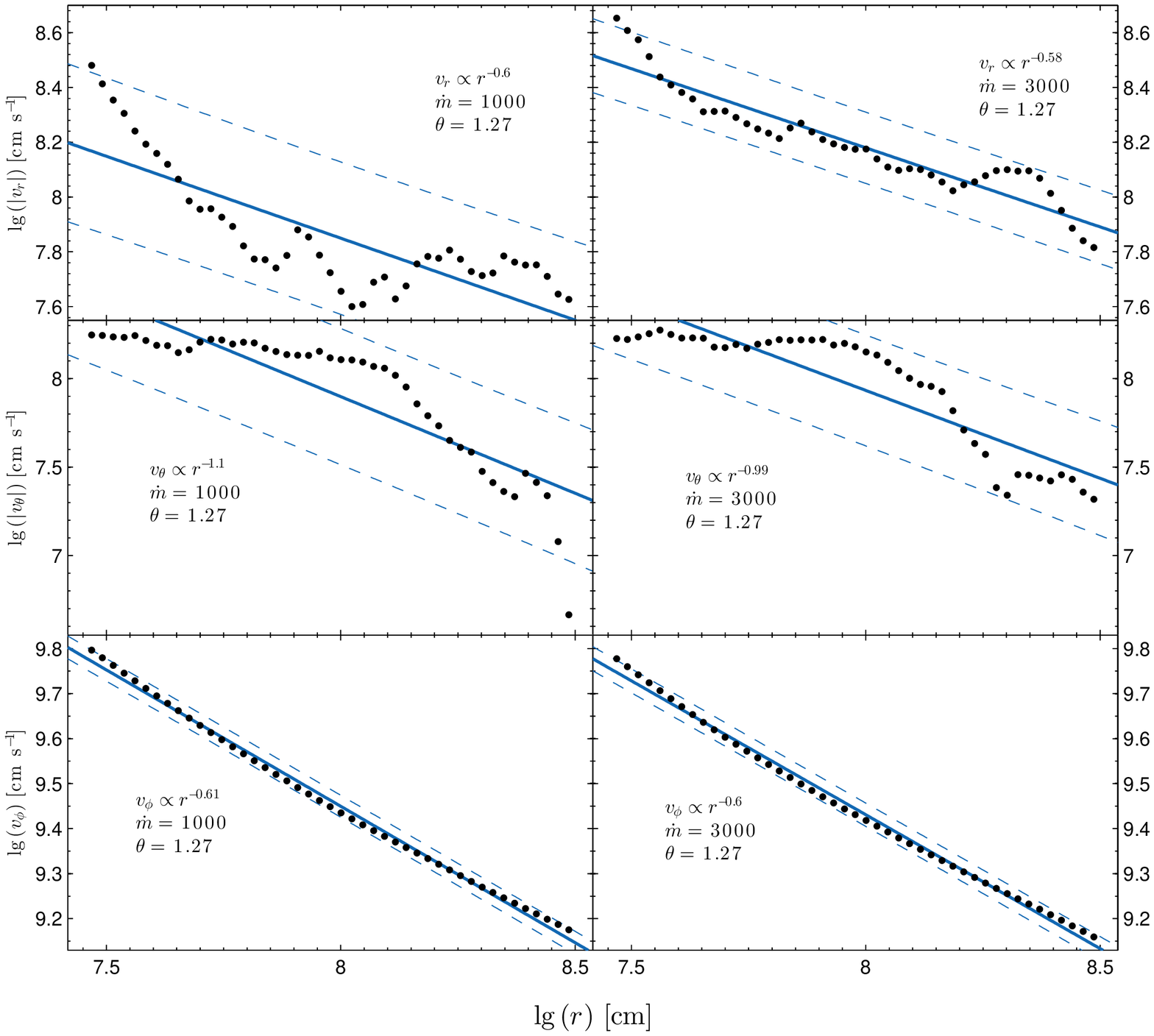}
 \caption{
  {The linear fit curves of $\lg{|v_r|}$, $\lg{|v_\theta|}$ and $\lg{v_\phi}$ at polar angle $\theta=1.2732$ rad in the inflow region. The black dots correspond to the simulation data, the solid lines correspond to the linear fitting, and the dashed lines correspond to the 95\% confidence bounds. For $\dot{m}=1000$ and $\dot{m}=3000$, the $R^2$ of $\lg{|v_r|}$ fits are 0.6448 and 0.8857, respectively, and the fitted slopes of $\lg{|v_r|}$ are -0.5981  (-0.7346, -0.4616) and -0.5775  (-0.6413, -0.5137), respectively, with 95\% confidence bounds in brackets; the $R^2$ of $\lg{|v_\theta|}$ fits are 0.7597 and 0.7979, respectively, and the fitted slopes of $\lg{|v_\theta|}$ are -1.093  (-1.281, -0.9035) and -0.9937  (-1.147, -0.8399), respectively; the $R^2$ of $\lg{v_\phi}$ fits are 0.9959 and 0.9952, respectively, and the fitted slopes of $\lg{v_\phi}$ are -0.6061  (-0.6181, -0.5941) and -0.5951  (-0.6078, -0.5824), respectively.}
 }
\end{figure}

\begin{figure}[htbp]
\plotone{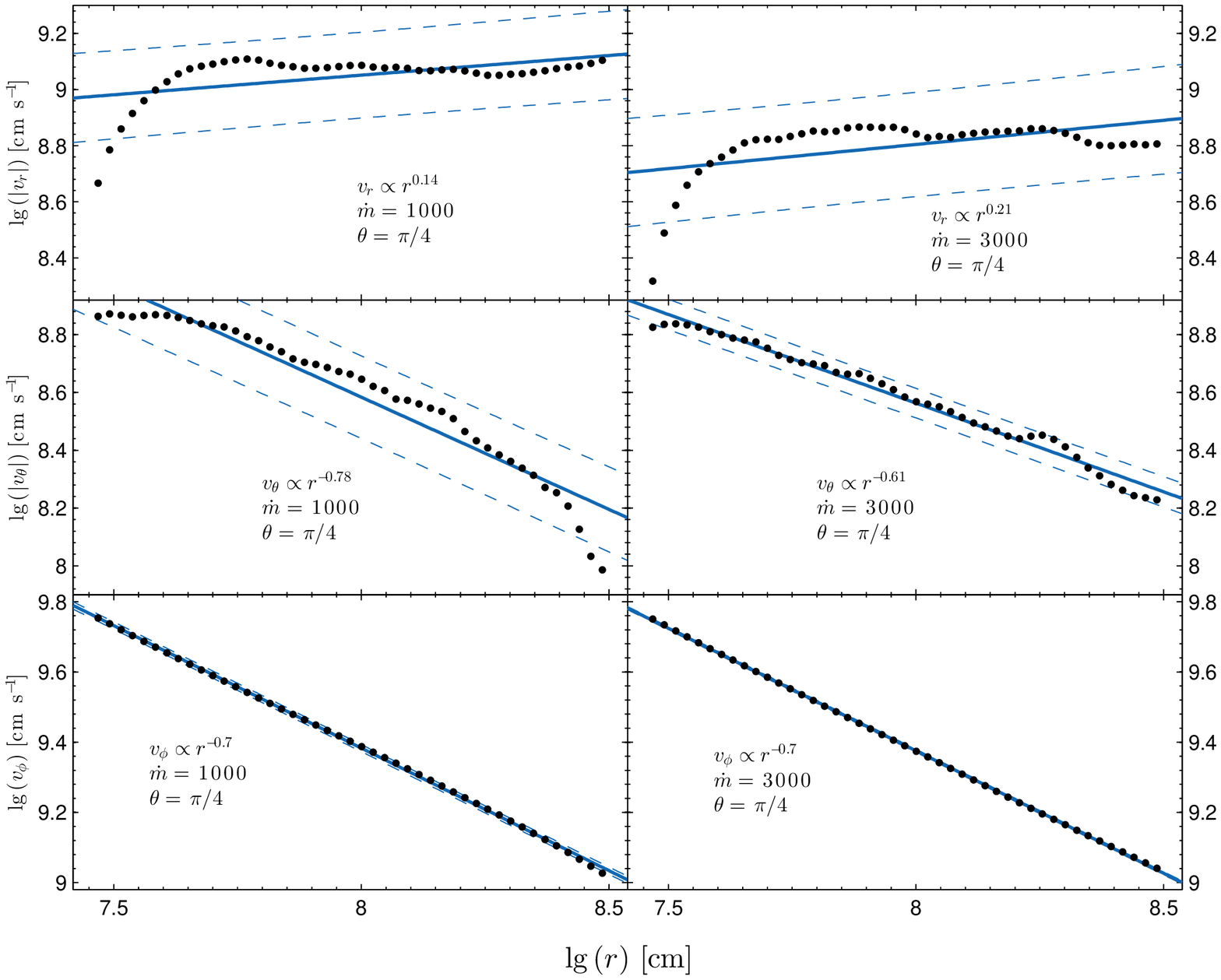}
 \caption{
 The linear fit curves of $\lg{|v_r|}$, $\lg{|v_\theta|}$ and $\lg{v_\phi}$ at polar angle $\theta=\pi/4$ in the outflow region. The black dots correspond to the simulation data, the solid lines correspond to the linear fitting, and the dashed lines correspond to the 95\% confidence bounds. For $\dot{m}=1000$ and $\dot{m}=3000$,  {the $R^2$ of $\lg{|v_r|}$ fits are 0.2478 and 0.3163, respectively, and the fitted slopes of $\lg{|v_r|}$ are 0.1399  (0.06495, 0.2149) and 0.2064  (0.1131, 0.2998), respectively,} with 95\% confidence bounds in brackets; the $R^2$ of $\lg{|v_\theta|}$ fits are 0.9209 and 0.982, respectively, and the fitted slopes of $\lg{|v_\theta|}$ are -0.7764  (-0.8464, -0.7064) and -0.6071  (-0.6324, -0.5818), respectively; the $R^2$ of $\lg{v_\phi}$ fits are 0.9995 and 0.9999, respectively, and the fitted slopes of $\lg{v_\phi}$ are -0.698  (-0.7029, -0.6931) and -0.698  (-0.6999, -0.6961), respectively.
 }
\end{figure}

\begin{figure}[htbp]
\plotone{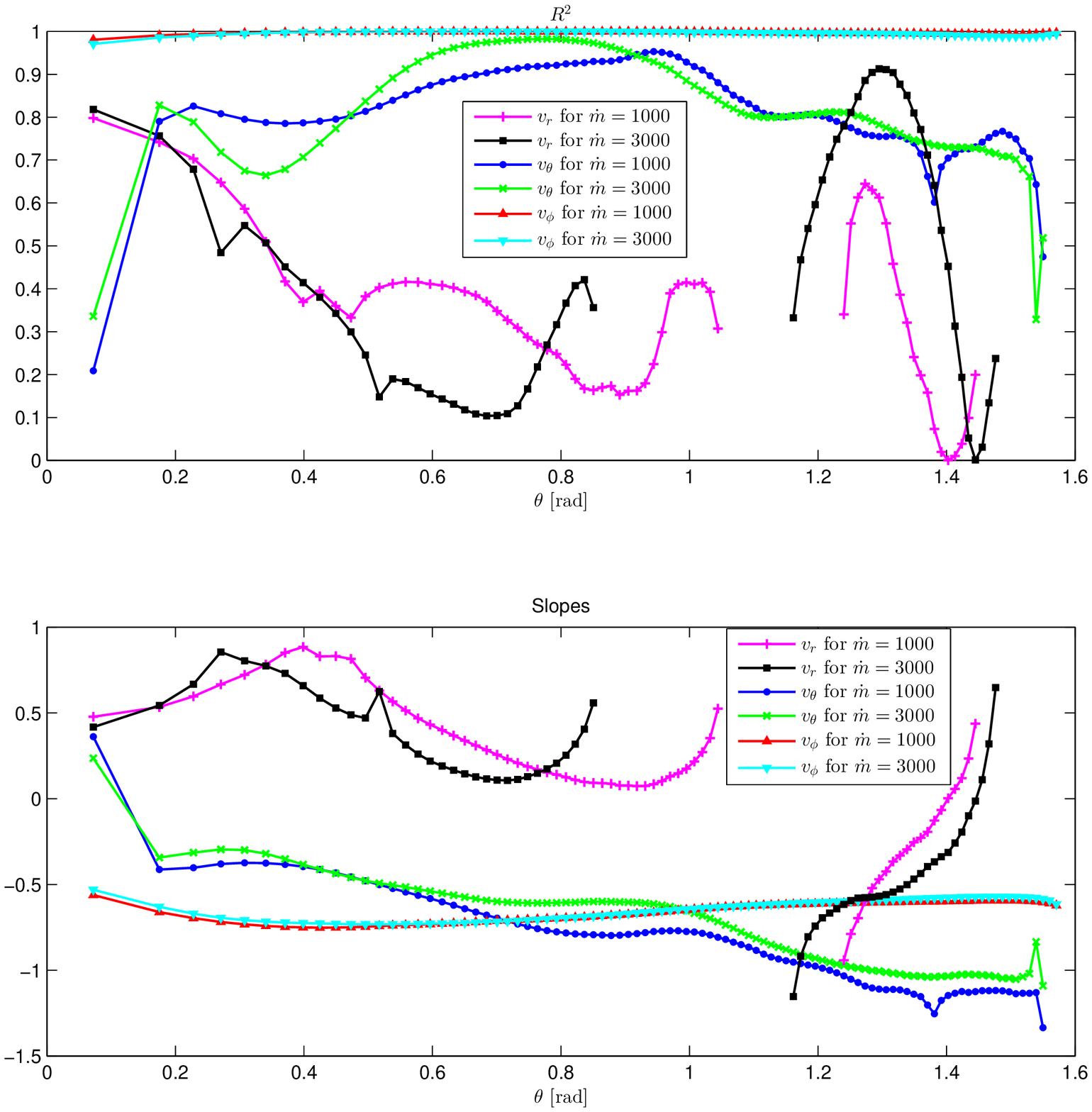}
\caption{
The fitted slopes and $R^2$ values of the linear fits of $\lg{|v_r|}$, $\lg{|v_\theta|}$ and $\lg{v_\phi}$ at different polar angles. The upper panel corresponds to the $R^2$ values, and the lower panel corresponds to the fitted slopes. Different markers represent different $\dot{m}$ and velocity components, as shown in the legend.  {For the fits of $\lg{|v_r|}$, we ignore $\theta$ values for which there exists $v_r=0$.}
}
\end{figure}

 {The fits of $\lg{|v_r|}$ displays two different types of behavior in the inflow and outflow regions, as shown in Figures 1 and 2. In the inflow region, the value of $v_r$ generally increases as radius decreases, but is strongly affected by the circulation patterns. Even after we remove the fits of $\theta$ values where $v_r$ changes sign due to circulation patterns (i.e. the worst cases), the average $R^2$ in the inflow region is only 0.31 and 0.59 for $\dot{m}=1000$ and $\dot{m}=3000$, respectively, with the best $R^2$ valules of 0.64 and 0.91, respectively, as shown in Table 1. In the outflow region, the value of $v_r$ first increases as radius decreases, then at some point it starts to decrease, as shown in the top panels of Figure 2, so that the fitted slopes of $\lg{|v_r|}$ are mostly positive in the outflow region in Figure 3. This is because when the accretion flow gets close to the black hole, stronger gravity and relativistic effects will gradually turn outflow into inflow (e.g. Figure 6 of OMNM05, in which inflow accretion rate becomes almost constant close to the black hole). Similar effects are also found in simulation works of ADAFs (Narayan et al. 2012; Yuan et al. 2012a), in which the inflow accretion rate are reported to be almost constant inside 10$r_g$. This means that in the outflow region near the black hole, as radius decreases, $v_r$ will decrease and eventually become negative, which corresponds to inflow. While this change usually happens inside 10$r_g$, which we ignore in the linear regression, it does have influence on the profile of $v_r$ in the radial direction, so that in the outflow region $v_r$ will start decreasing from some radius as $r$ gets closer to 10$r_g$.}

The fits of $\lg{v_\phi}$ are good, with fitted $R^2$ very close to 1  {(0.99 on average for both $\dot{m}=1000$ and $\dot{m}=3000$)} and fitted slopes generally resembling the same value. So the self-similar model  {describes the radial distribution of $v_\phi$ well}. The average fitted slopes of $\lg{v_\phi}$ are -0.65 and -0.64 for $\dot{m}=1000$ and $\dot{m}=3000$, respectively, which differ from the slope of -0.5 in self-similar assumptions we use, but are still acceptable.  {The fits of $\lg{|v_\theta|}$ are also good for most values of $\theta$, with an average $R^2$ of 0.81 and 0.82 for $\dot{m}=1000$ and $\dot{m}=3000$, respectively. The bad fits of $\lg{|v_\theta|}$ appear near the equatorial plane and the polar axis, which is not surprising as in these places $v_\theta$ gets close to 0 due to symmetry and is thus much influenced by numerical errors. The fits of $\lg{|v_\theta|}$ are generally worse than those of $\lg{v_\phi}$, as the profiles of $v_\theta$ are influenced by the circulation patterns we mentioned above, while those of $v_\phi$ are not.} The slopes of the fits of $\lg{v_\phi}$ and $\lg{|v_\theta|}$ are slightly different from -0.5, which means that their radial distributions are not strictly proportional to that of the corresponding Keplerian velocity.

 {It should be noted that, near the equatorial plane close to the black hole, for both $\dot{m}=1000$ and $\dot{m}=3000$, there is a small region which is severely influenced by the circulation patterns and displays another contour of $v_r=0$, aside from the boundary between inflow and outflow region (c.f. the velocity field plot Figure 11 in Section 4). Although we ignore the $\theta$ values for which there exists $v_r=0$, near this region the $v_r$ profiles are still much affected and the values of $v_r$ drops significantly, making some fits of $\lg{|v_r|}$ in the inflow region bad (with bad $R^2$ values and positive slopes, as shown in Figure 3 and Table 1). If we regard the circulation patterns as perturbation caused by convection over a self-similar configuration, then it appears that the effect is much stronger in $r$ direction than in $\theta$ direction in spherical coordinates.}

\subsection{Fitting density $\rho$}
The fitting results of density $\rho$ are presented in this subsection. Figure 4 displays the linear fitting results of $\lg{\rho}$ at $\theta=\pi/2$ (i.e. on the equatorial plane) and $\theta=\pi/4$ for $\dot{m}=1000$ and $\dot{m}=3000$, respectively. All the physical quantities are in cgs units. The fitting results are good  {with $R^2$ values of 0.9591 and 0.9347 at $\theta=\pi/2$, and 0.8902 and 0.9209 at $\theta=\pi/4$, for $\dot{m}=1000$ and $\dot{m}=3000$, respectively}.
\begin{figure}[htbp]
\begin{center}
 \includegraphics[width=5 in]{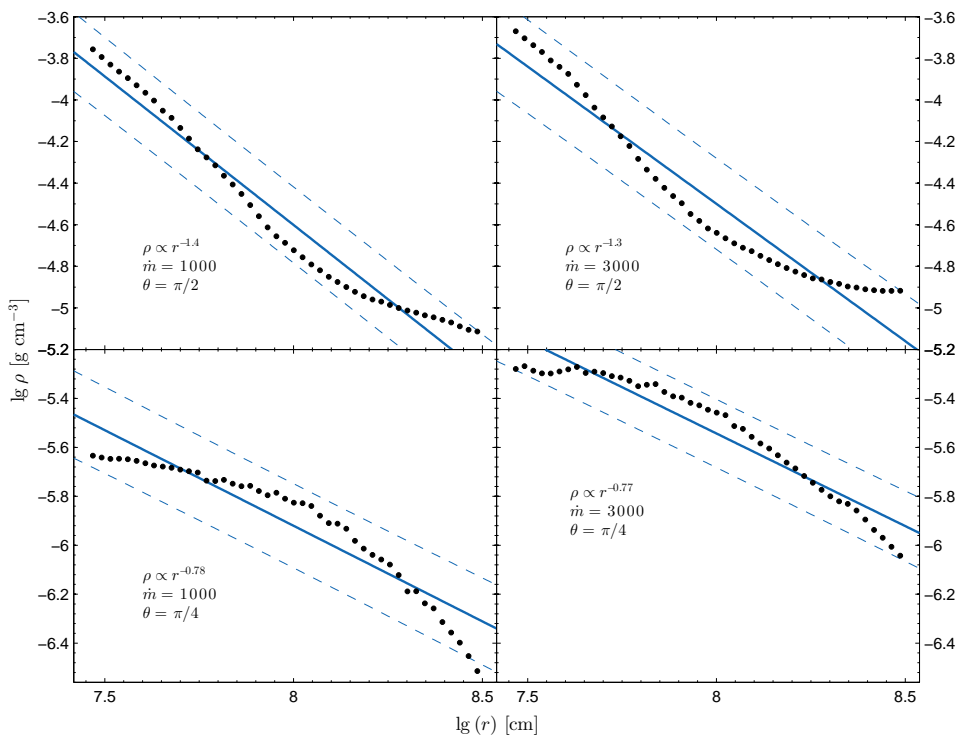}
 \caption{The fitting results of $\lg{\rho}$. The black dots correspond to the simulation data, the solid lines correspond to the linear fitting, and the dashed lines correspond to the 95\% confidence bounds. For $\dot{m}=1000$, the $R^2$ of $\lg{\rho}$ fits are 0.9591 at $\theta=\pi/2$, and 0.8902 at $\theta=\pi/4$, respectively, and the corresponding fitted slopes are -1.428 (-1.518, -1.338) and  -0.7812 (-0.8656, -0.6968), respectively, with 95\% confidence bounds in brackets.  For $\dot{m}=3000$, the $R^2$ are 0.9347 at $\theta=\pi/2$, and 0.9209 at $\theta=\pi/4$, and the corresponding fitted slopes are  -1.313  (-1.419, -1.206) and -0.7655 (-0.8345, -0.6965), respectively.}
   \label{fig1}
\end{center}
\end{figure}

Figure 5 gives a general view of the fitting results of density $\rho$ at different polar angles. The upper panel corresponds to $R^2$ values, which are close to 1  {(0.94 on average for both $\dot{m}=1000$ and $\dot{m}=3000$)}, so the fitting results are good. The lower panel corresponds to the fitted slopes at different polar angles, which generally resemble the same value on each curve, so the self-similar model  {describes the radial distribution of density in the the simulation data well}. The average slope here is -0.85 for $\dot{m}=1000$, and -0.74 for $\dot{m}=3000$, which correspond to $n \thicksim 0.85$ and $n \thicksim 0.74$, respectively, according to Eq.(12).
\begin{figure}[htbp]
\plotone{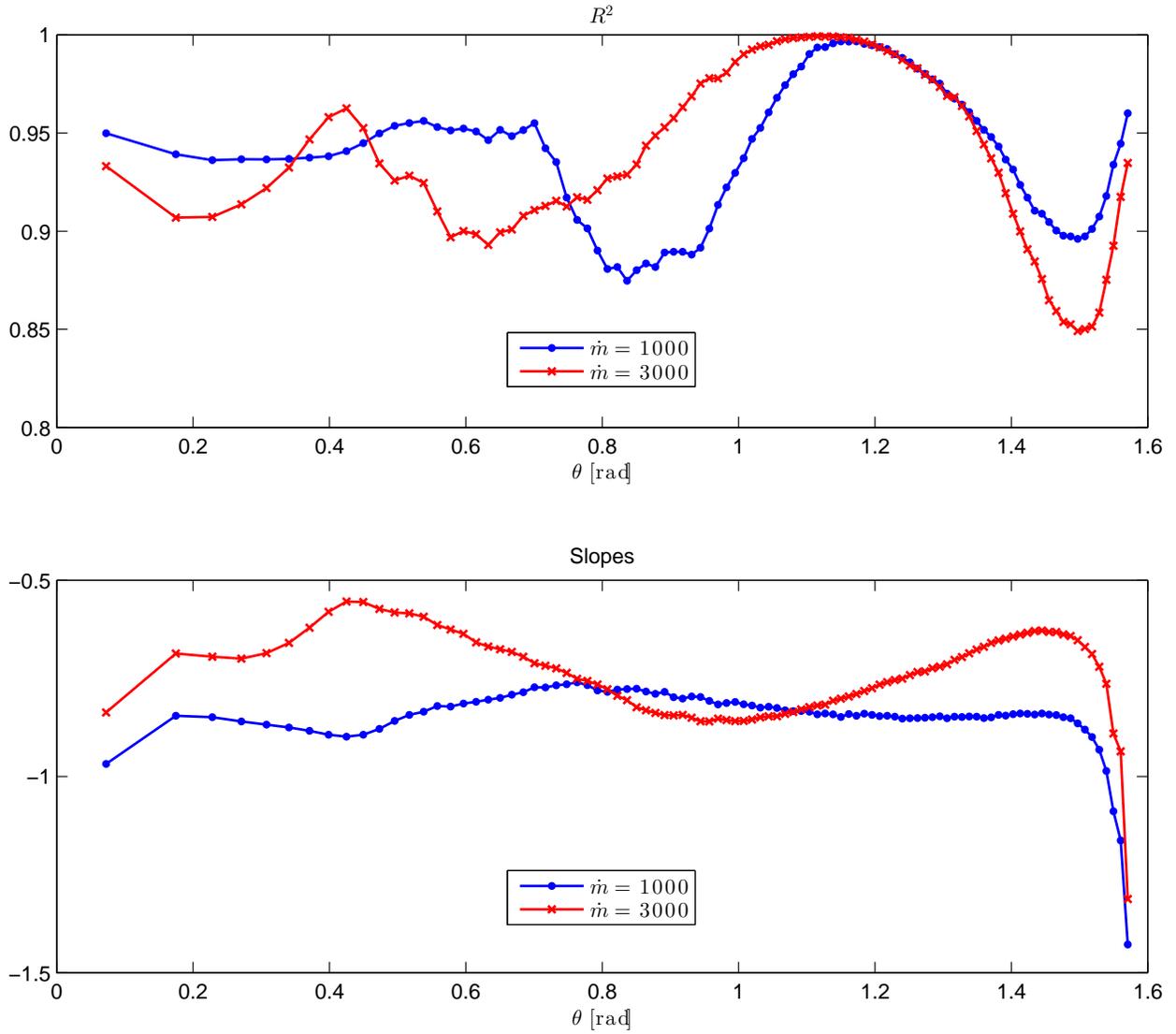}
\caption{
The fitted slopes and $R^2$ values of the linear fits of $\lg{\rho}$ at different polar angles. The upper panel corresponds to the $R^2$ values, and the lower panel corresponds to the fitted slopes. Different markers correspond to different $\dot{m}$, as shown in the legend.
}
\end{figure}

\subsection{Fitting pressure $p$}
The total pressure in the simulations comes from both gas pressure and radiation pressure. The radiation stress tensor is almost isotropic in the optically thick region, so we have
\begin{equation}\label{}
    p_\mathrm{rad}=E_\mathrm{rad}/3,
\end{equation}
in which $E_\mathrm{rad}$ is the radiation energy density (per unit volume). Note that for the simulation results, in the region very close to the polar axis, the optical depth could be smaller than 1.
The optical depth $\tau$ can be approximated by $\cal R$ (Kato et al. 2008):
\begin{equation}\label{}
    {\cal R}=\frac{\left| \nabla E \right|}{(\kappa_\mathrm{abs}+\kappa_\mathrm{sca})\rho E} \sim \frac{1}{\tau}
\end{equation}
Figure 6 shows the contour plots of $\cal R$ in the converged regions of the simulations for $\dot{m}=1000$ and $\dot{m}=3000$. It can be seen that most of the space is occupied by optically thick accretion flows for both simulations. The optically thin regions are very small, and only have negligible influence here.
\begin{figure}[htbp]
  \centering
\begin{minipage}[c]{0.5\textwidth}
\centering
  \includegraphics[width=6.5cm]{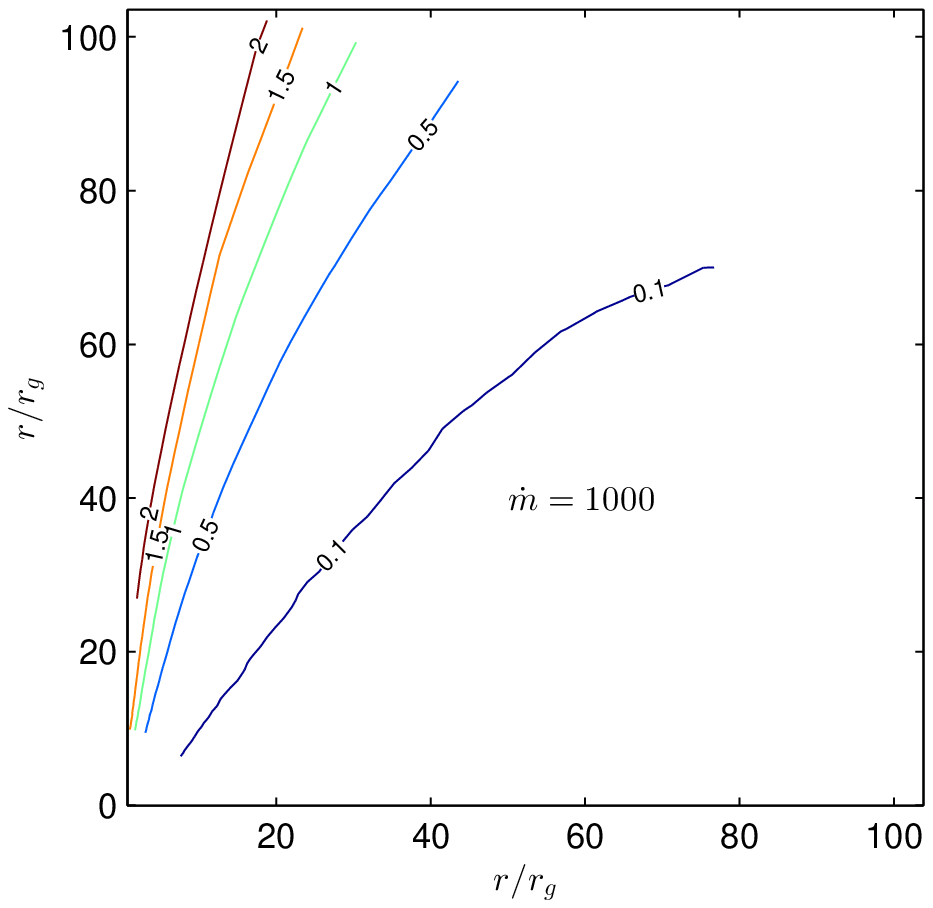}
\end{minipage}%
\begin{minipage}[c]{0.5\textwidth}
\centering
  \includegraphics[width=6.5cm]{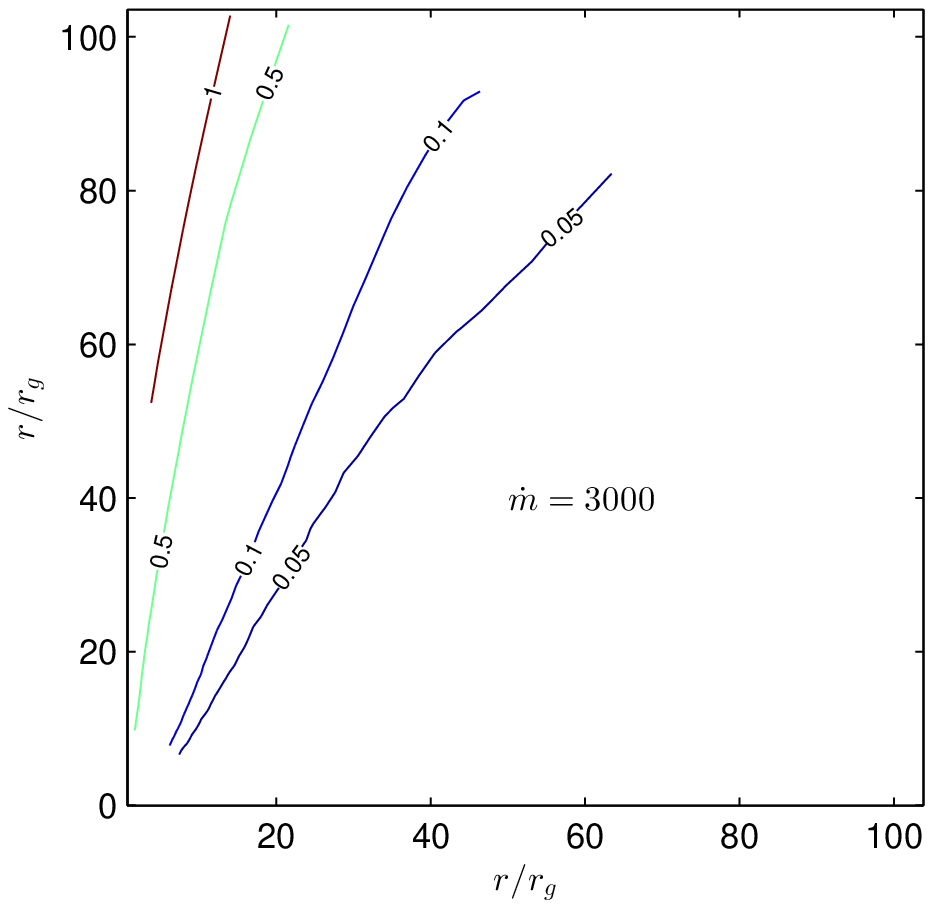}
\end{minipage}%
\caption{Contour plots of $\cal R$ in the converged regions of the simulation data. $\cal R \sim$ $1/\tau$ indicates the local optical depth. The left panel is for the simulation with $\dot{m}=1000$, while the right panel is for the simulation with $\dot{m}=3000$.}
\end{figure}
So the total pressure can be written as
\begin{equation}\label{}
    p=p_\mathrm{rad}+p_\mathrm{gas}=E_\mathrm{rad}/3+(\gamma-1)E_\mathrm{gas},
\end{equation}
in which $\gamma$ is the heat capacity ratio which is taken as 5/3 in both simulations, and $E_\mathrm{gas}$ is the gas internal energy density (per unit volume).

Figure 7 displays the linear fitting results of $\lg{p}$ at $\theta=\pi/2$ (i.e. on the equatorial plane) and $\theta=\pi/4$ for $\dot{m}=1000$ and $\dot{m}=3000$, respectively. All the physical quantities are in cgs units. The fitting results are  {good with $R^2$ values of 0.9997 and 0.9997 at $\theta=\pi/2$, and 0.9865 and 0.9895 at $\theta=\pi/4$, for $\dot{m}=1000$ and $\dot{m}=3000$ respectively}.
\begin{figure}[htbp]
\begin{center}
 \includegraphics[width=5 in]{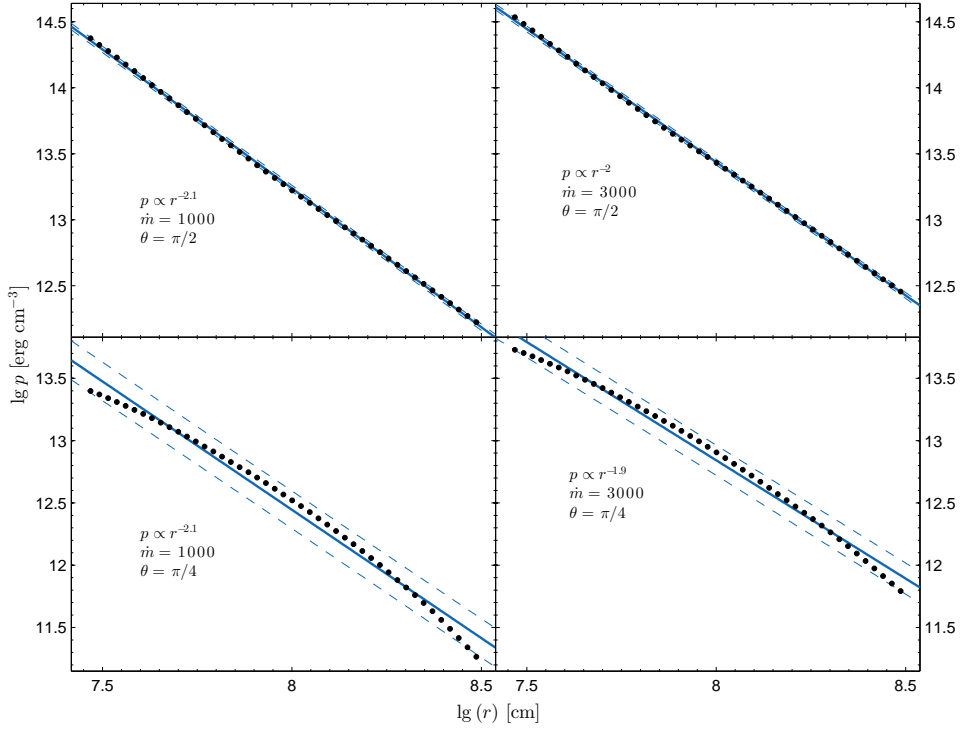}
 \caption{The fitting results of $\lg{p}$. The black dots correspond to the simulation data, the solid lines correspond to the linear fitting, and the dashed lines correspond to the 95\% confidence bounds. For $\dot{m}=1000$, the $R^2$ of $\lg{p}$ fits are 0.9997 at $\theta=\pi/2$, and 0.9865 at $\theta=\pi/4$, respectively, and the corresponding fitted slopes are -2.105 (-2.116, -2.093) and -2.06 (-2.134, -1.985), respectively, with 95\% confidence bounds in brackets.  For $\dot{m}=3000$, the $R^2$ are 0.9997 at $\theta=\pi/2$, and 0.9895 at $\theta=\pi/4$, and the corresponding fitted slopes are  -2.018 (-2.028, -2.007) and -1.901  (-1.961, -1.841), respectively.}
   \label{fig1}
\end{center}
\end{figure}

Figure 8 gives a general view of the fitting results of total pressure $p$ at different polar angles. The upper panel corresponds to $R^2$ values, which are very close to 1  {(0.99 on average for both $\dot{m}=1000$ and $\dot{m}=3000$)}, so the fitting results are good. The lower panel corresponds to the fitted slopes at different polar angles, which generally resemble the same value on each curve, so the self-similar model  {describes the radial distribution of total pressure in the simulation data well}. The average slope here is -2.03 for $\dot{m}=1000$, and -1.91 for $\dot{m}=3000$, which corresponds to $n \thicksim 1.03$ and $n \thicksim 0.91$, respectively, according to Eq.(13). Note that the values of $n$ obtained here {do} not agree completely with the $n$ values obtained from the fits of $\rho$, but are quite close. This will be discussed in detail later in the paper.
\begin{figure}[htbp]
\plotone{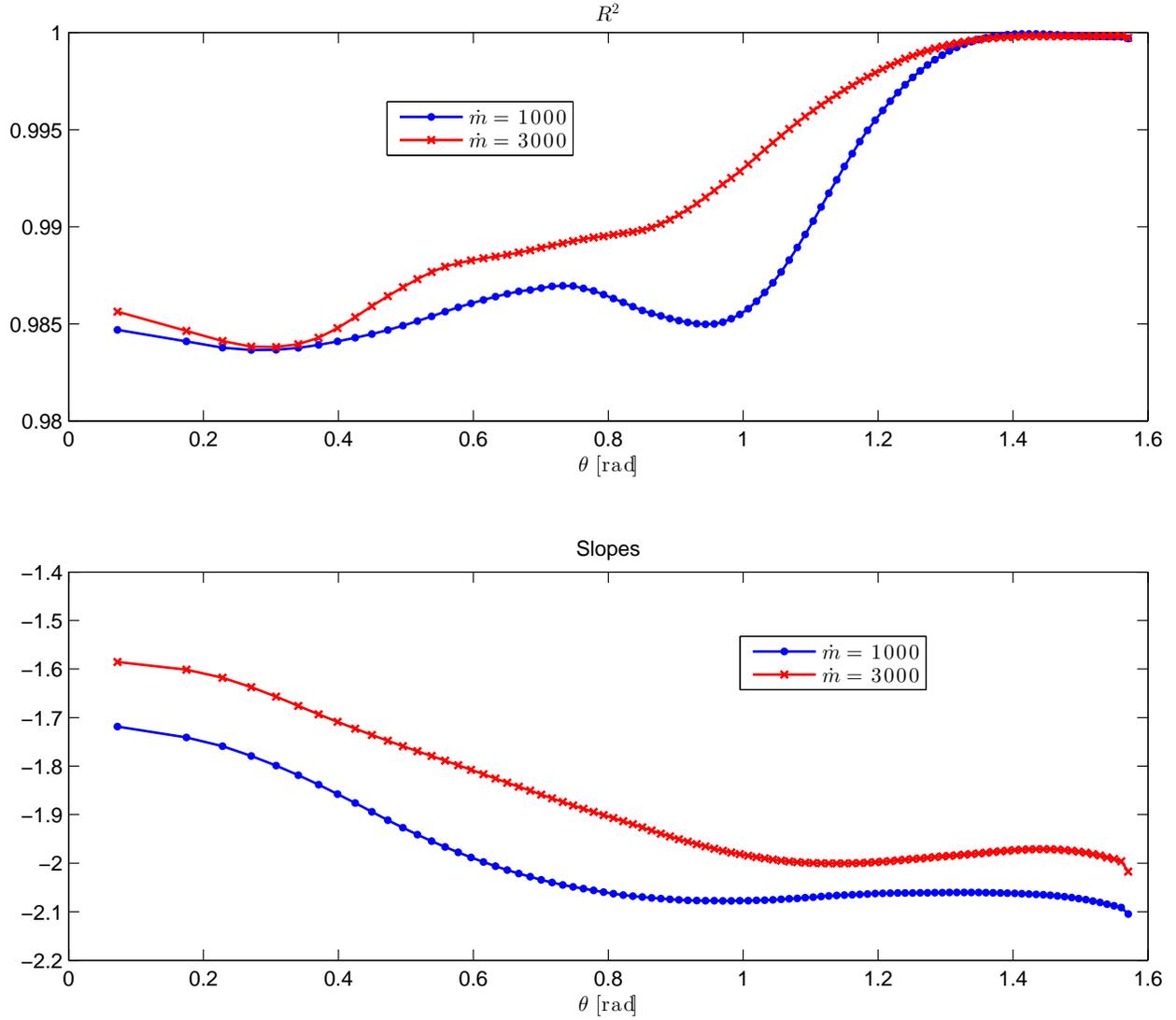}
\caption{
The fitted slopes and $R^2$ values of the linear fits of $\lg{p}$ at different polar angles. The solid curves correspond to the $R^2$ values, and the dashed curves correspond to the fitted slopes. Different markers correspond to different $\dot{m}$, as shown in the legend.
}
\end{figure}
\section{Comparing the steady model solutions with simulation results}
In this section we compare the steady accretion disk model solutions with the simulation results. First we determine the input parameters for the steady model from the simulation data. Then the steady model solutions corresponding to these parameters are calculated and compared with the simulation results.
\subsection{Determining input parameters}
Four input parameters are required for the steady accretion disk model, namely ($\alpha$,$f$,$\gamma_\mathrm{equ}$,$n$). The simulation data are calculated with $\alpha=0.1$ and $\gamma=5/3$ as mentioned in section 2. According to Eq.(7), $\gamma_\mathrm{equ}$ is determined by both $\gamma$ and $\beta$. Figure 9 displays the contour maps of gas pressure ratio $\beta$ obtained from the simulation data. It can be seen that, in the converged region of the simulations, $\beta$ is generally below 0.2, except for a very small region close to 10$r_g$. The converged regions in both simulations are radiation-pressure dominated, with a maximum $\beta_\mathrm{max}= 0.386$ (corresponding to $\gamma_\mathrm{equ}=1.41$) for $\dot{m}=1000$, and $\beta_\mathrm{max}= 0.344$ (corresponding to $\gamma_\mathrm{equ}=1.40$) for $\dot{m}=3000$. So here we can safely take $\gamma_\mathrm{equ}=4/3$, which corresponds to extremely radiation-pressure dominated accretion flows.
\begin{figure}[htbp]
  \centering
\begin{minipage}[c]{0.5\textwidth}
\centering
  \includegraphics[width=6.5cm]{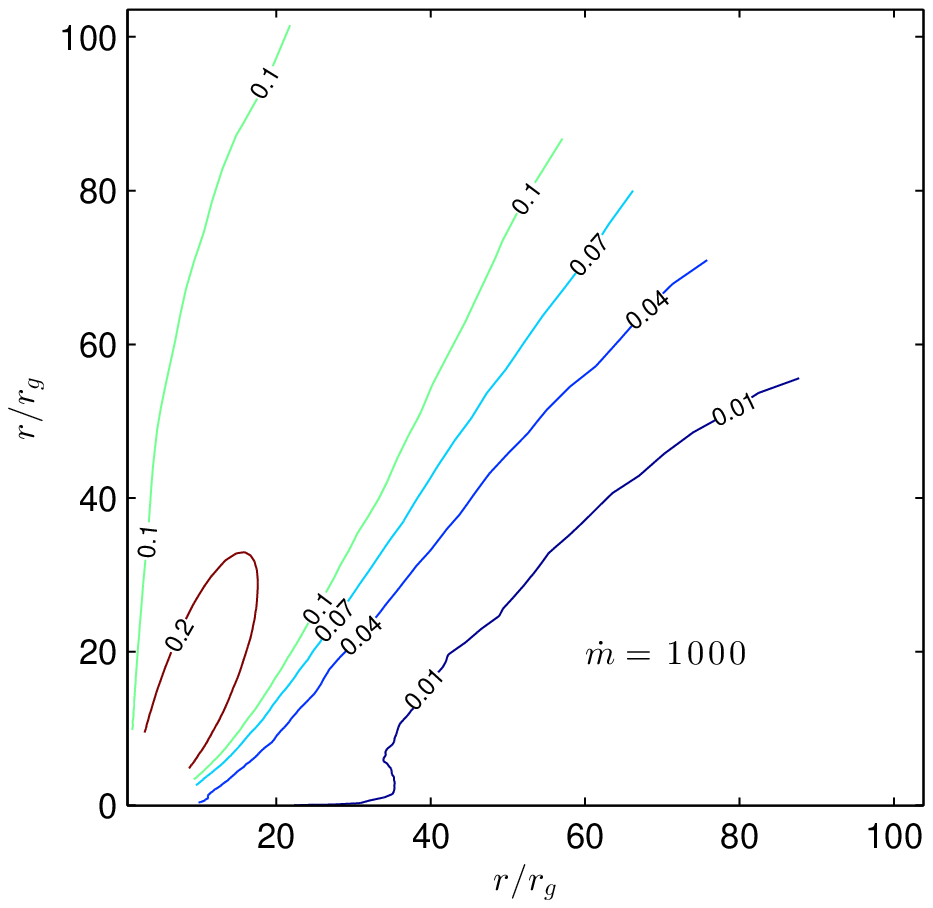}
\end{minipage}%
\begin{minipage}[c]{0.5\textwidth}
\centering
  \includegraphics[width=6.5cm]{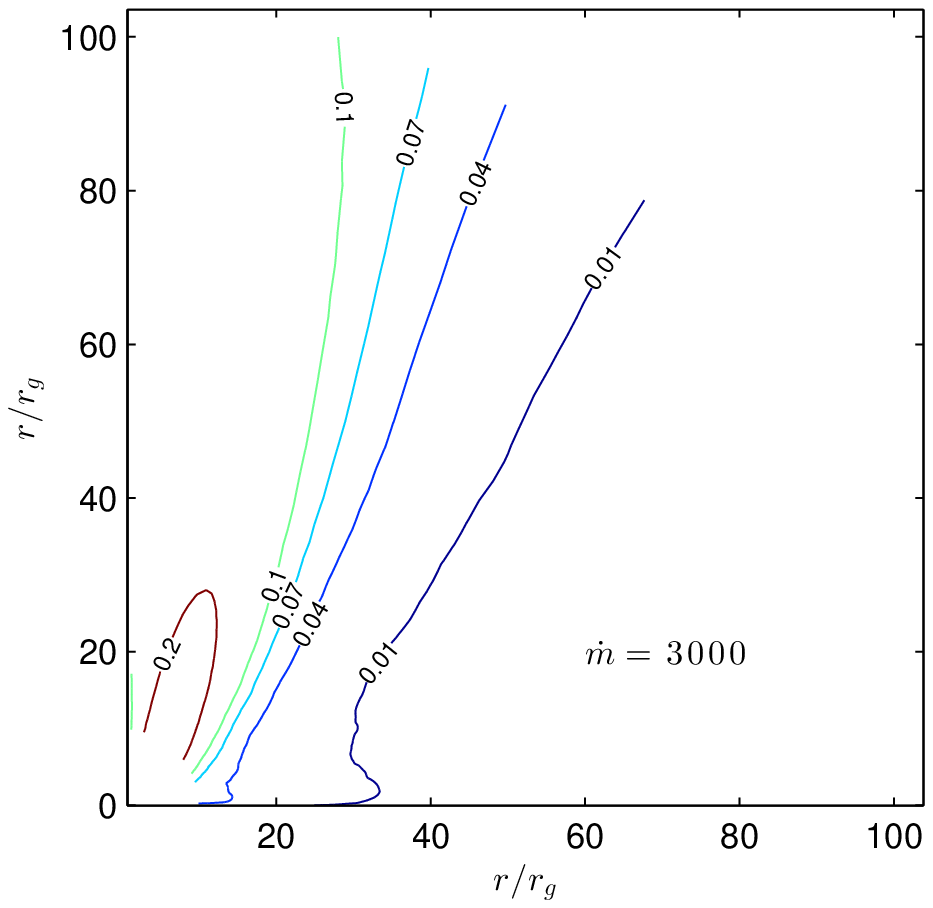}
\end{minipage}%
\caption{Contour plots of gas pressure ratio $\beta$ in the converged regions of the simulation data. The left panel is for the simulation with $\dot{m}=1000$, while the right panel is for the simulation with $\dot{m}=3000$.}
\end{figure}

The parameter $n$ can be obtained from either the fits of density $\rho$ or the fits of total pressure $p$, as indicated by Eqs.(12) and (13). As discussed above, we get $n \thicksim 0.85$ and $n \thicksim 1.03$, from the fits of $\rho$ and $p$, respectively, for $\dot{m}=1000$. For $\dot{m}=3000$, we get $n \thicksim 0.74$ and $n \thicksim 0.91$, respectively. However, the relation indicated by Eq.(13) is based on the assumption:
\begin{equation}\label{}
    c_s^2 \propto \frac{GM}{r},
\end{equation}
in which $c_s$ is the sound speed. As discussed in section 2, the profile of $v_\phi$ actually scales in the radial direction with $r^{-0.65}$ on average for $\dot{m}=1000$, and $r^{-0.64}$ on average for $\dot{m}=3000$. It is possible that $c_s$ also deviates from the $r^{-0.5}$ scaling in the radial direction. So it is better to determine $n$ with the density profiles. Here we set $n=0.85$ for $\dot{m}=1000$, and $n=0.74$ for $\dot{m}=3000$, from the fits of density $\rho$. It does not differ a lot from the values obtained from the fits of pressure $p$, anyway.

The energy equations adopted in the simulations are
\begin{equation}\label{}
    \frac{\partial E_\mathrm{gas}}{\partial t}+\bm{\nabla}\cdot(E_\mathrm{gas}\bm{v})+p_{gas}\bm{\nabla}\cdot\bm{v}=q^+-\rho j+\rho c\kappa_\mathrm{abs}E_\mathrm{rad},
\end{equation}
\begin{equation}\label{}
    \frac{\partial E_\mathrm{rad}}{\partial t}+\bm{\nabla}\cdot(E_\mathrm{rad}\bm{v})+\bm{\nabla}\bm{v}:\bm {\rm{P}} = -\bm{\nabla}\cdot\bm{F}+\rho j-\rho c\kappa_\mathrm{abs}E_\mathrm{rad},
\end{equation}
in which $\bm{v}$ is the velocity, $\bm{P}$ is the radiation pressure tensor, $q^+$ is the viscous heating rate, $j$ is the emissivity per unit mass, and $\kappa_\mathrm{abs}$ is the absorption opacity. Eq.(19) is the energy equation of the gas, and Eq.(20) is the energy equation of the radiation. The left hand side of these two equations, excluding the time derivative terms, are the advection of the gas internal energy and the advection of the radiation field, respectively. In the steady model, the advection term $q_\mathrm{adv}$ includes both sources of advection. Correspondingly, if we sum up Eq.(19) and (20), we can get
\begin{equation}\label{}
    \frac{\partial (E_\mathrm{gas}+E_\mathrm{rad})}{\partial t}+q_\mathrm{adv}=q^+-\bm{\nabla}\cdot\bm{F}.
\end{equation}
$\bm{\nabla}\cdot\bm{F}$ is similar to the radiation cooling in one-dimensional accretion disk models. It represents the change of the radiation energy in a fixed region due to the transportation via the radiative flux $\bm{F}$. The advective factor $f$ can be calculated from Eq.(21):
\begin{equation}\label{}
    f \equiv \frac{q_\mathrm{adv}}{q^+}=1-\frac{\frac{\partial (E_\mathrm{gas}+E_\mathrm{rad})}{\partial t}+\bm{\nabla}\cdot\bm{F}}{q^+}.
\end{equation}
As we focus on the converged region in the simulation which has achieved quasi-steady state, the time derivative terms are typically much smaller than other terms ($10^{-3} - 10^{-4}$ times $q^+$),  {so the values of $f$ mainly depend on $q^+$ and $\bm{\nabla}\cdot\bm{F}$. We then average $f$ over time and radial direction, to get its profile in $\theta$ direction. Figure 10 displays the result.} Basically $f$ is close to 1 on the equatorial plane (where $\theta=\pi/2$), decreases as inclination decreases, and then increases again as $\theta$ gets close to the polar axis, and values of $f$ can get very large (greater than 10, not shown in Figure 10 to display more details for lower values of $f$) near the polar axis.
This can be explained by the photon trapping effect (Begelman 1978; Takeuchi \& Mineshige 2009; also see OMNM05) and the existence of strong outflow near the polar axis. The photons generated near the equatorial plane are more effectively trapped than those generated at higher latitude; as $\theta$ decreases, it is easier for the generated photons to escape the accretion flow, so $f$ decreases accordingly.  {However, below a certain value of $\theta$, more photons are generated in the accretion flow at lower latitude and carried inside by the radiative flux than those escaping to higher latitude}, so that $\bm{\nabla}\cdot\bm{F}$ becomes negative, which causes $f$ to become larger than 1. These photons are then carried outward by the strong outflow near the polar axis, together with kinetic and internal energy of the gas, in the form of $q_\mathrm{adv}$. This is also verified later as we investigate the kinetic energy flux carried by outflows in Section 4.4. It should be noted that the region near the polar axis is not described in the steady model, so it does not influence our calculation. The dashed lines in Figure 10 represent the boundary of $\theta$ in the steady model calculations.

The values of $f$ are comparably larger in the simulation for $\dot{m}=3000$ than those for $\dot{m}=1000$. It is natural because for supercritical accretion flows, larger accretion rate corresponds to heavier photon trapping, which causes $f$ to be larger.
\begin{figure}[htbp]
\plotone{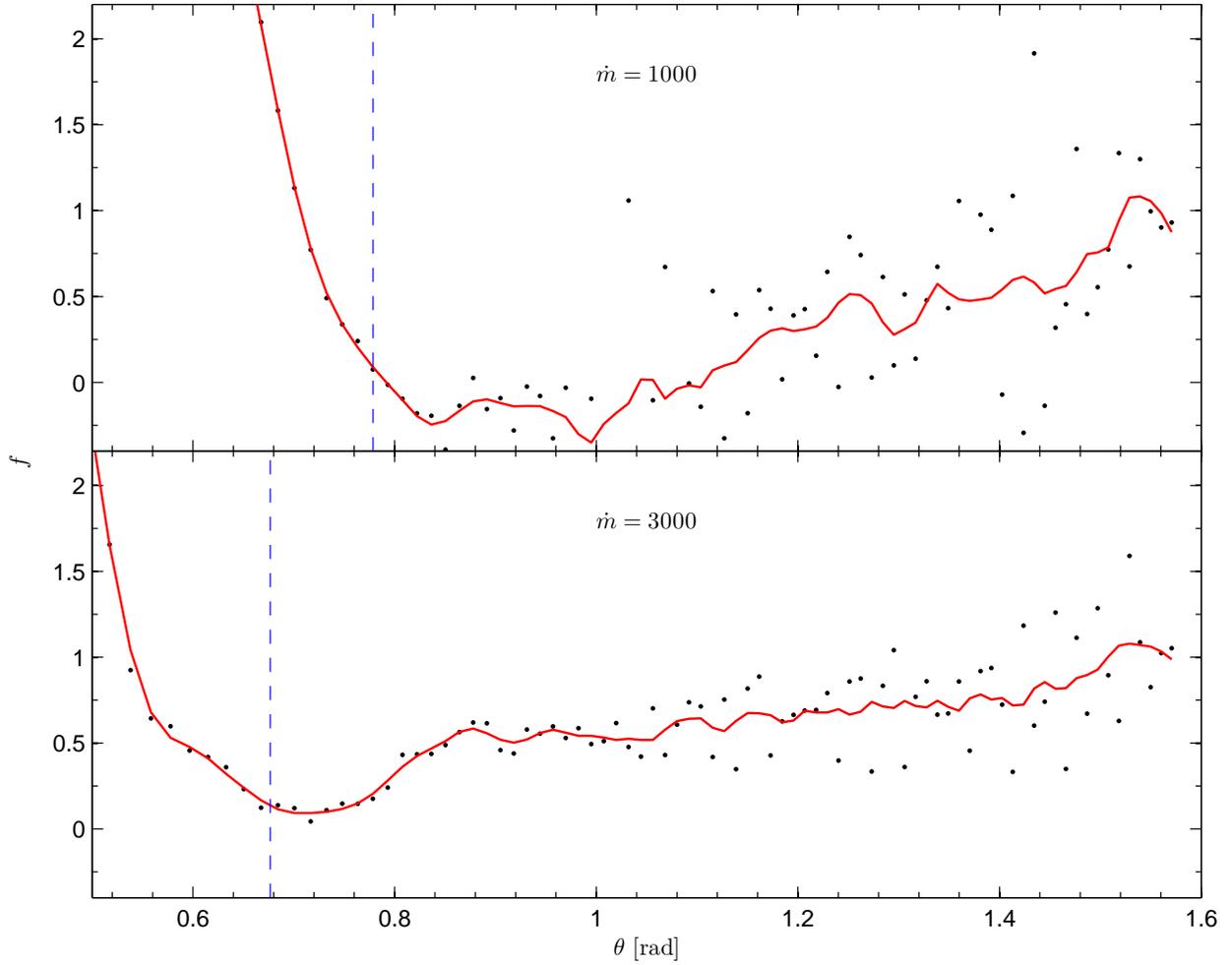}
\caption{
Profiles of the advective factor $f$ in $\theta$ direction,  {calculated by taking its average over time and radial direction}. Black dots correspond to the simulation data, the solid curves are data smoothed by a locally weighted linear least-squares regression method, and the dashed lines represent the boundary of the steady model calculations in $\theta$ direction.
}
\end{figure}

\subsection{Comparing the velocity fields}
In this subsection, we present the velocity field plots of the simulation results and the steady model solutions, as shown in Figure 11. The solid curves in each panel correspond to the surface of the inflow region at which $v_r=0$. The dashed lines correspond to the upper boundary of the steady model calculations beyond which our self-similar model can no longer describe (see JW11 for more details). In each velocity field plot of the simulation data, there is a small region wrapped by a solid curve near the equatorial plane. That corresponds to a circular pattern of the accretion flow, which arises from convection, as mentioned in Section 3.
\begin{figure}[htbp]
\plotone{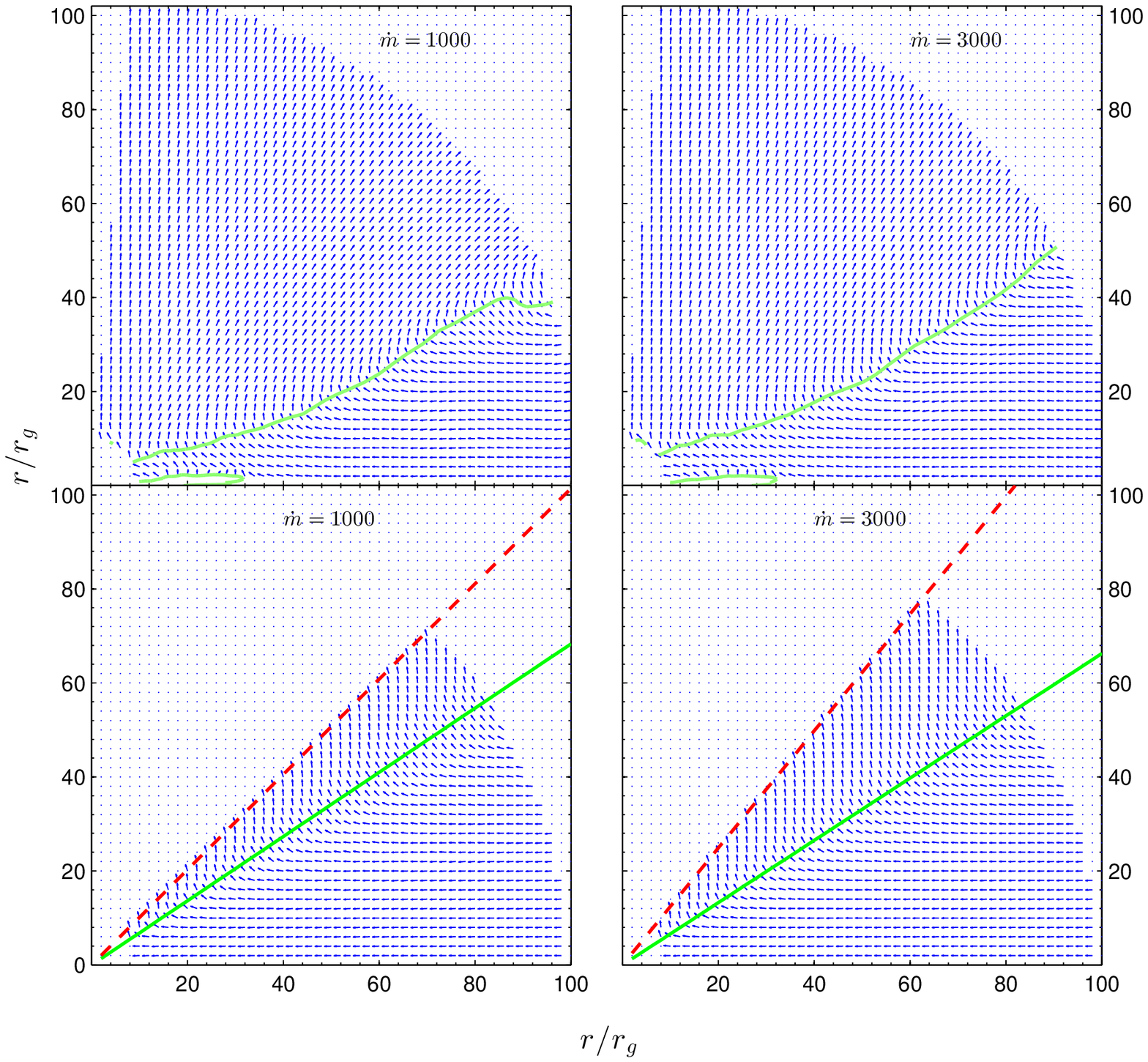}
\caption{
The velocity field plots of the simulation (upper panels) and steady model (lower panels) results. In each panel, the solid curves correspond to the surface of the inflow region, while the dashed line corresponds to the calculation limit of the steady models. The solid curves near the equatorial plane in the upper panels correspond to circular patterns of accretion flow in the simulation results, which arise from convection.
}
\end{figure}

It can be seen that for both accretion rates, the simulation result displays a smaller region of inflow than the steady model result. For $\dot{m}=1000$, the upper surface of the inflow region resides at $\theta=1.17$ rad ($\sim 67.0^\circ$) on average for the simulation, while it is at $\theta=0.972$ rad ($\sim 55.7^\circ$) for the steady model. For $\dot{m}=3000$, the upper surface of the inflow region resides at $\theta=1.06$ rad ($\sim 60.7^\circ$) on average for the simulation, while it is at $\theta=0.986$ rad ($\sim 56.5^\circ$) for the steady model. Besides that, in the outflow region of the simulation, the outward radial motion dominates the accretion flow, while in the steady model the flow follows a combination of $v_r$ and $v_\theta$. This indicates that either $v_r$ is much larger, or $v_\theta$ is much smaller, in the outflow region of the simulation results than those of the steady model results (in Section 4.3 we will see that $v_r$ is underestimated at high latitude in the steady model).

As the velocity field plots of the simulations here are based on time-averaged data, the convective motions in large circular pattern are not so obviously shown as in a snapshot of the simulation (cf. OMNM05). However, convection still works, though not dominant, as a transport mechanism of energy, mass and angular momentum. This will be discussed in Section 5.

\subsection{Comparing the variable profiles in $\theta$ direction}
In this subsection, we compare the profiles of velocities, density and pressure along the $\theta$ direction, obtained from the steady model solutions and the simulation data. In the steady model, the profiles are presented in the form of $v_r(\theta), v_\theta(\theta), v_\phi(\theta), \rho(\theta)$ and $p(\theta)$. The corresponding forms in the simulation results are
\begin{eqnarray}
  v_{r,\mathrm{sim}}(\theta) &=& \frac{v_{r,\mathrm{sim}}}{v_\mathrm{K}}, \\
  v_{\theta,\mathrm{sim}}(\theta) &=& \frac{v_{\theta,\mathrm{sim}}}{v_\mathrm{K}}, \\
  v_{\phi,\mathrm{sim}}(\theta) &=& \frac{v_{\phi,\mathrm{sim}}}{v_\mathrm{K}}, \\
  \rho_\mathrm{sim}(\theta) &=& \frac{\rho_\mathrm{sim}}{\rho_{\mathrm{sim},\theta=\pi/2}}, \\
  p_\mathrm{sim}(\theta) &=& \frac{p_\mathrm{sim}}{\rho_\mathrm{sim}v_\mathrm{K}^2} \rho_\mathrm{sim}(\theta)= \frac{p_\mathrm{sim}}{\rho_{\mathrm{sim},\theta=\pi/2}} \cdot \frac{1}{v_\mathrm{K}^2},
\end{eqnarray}
in which the subscript 'sim' indicates the data from simulation results and $v_\mathrm{K}$ is the Keplerian velocity
\begin{equation}\label{}
    v_\mathrm{K}=\sqrt{\frac{GM}{r-r_g}}.
\end{equation}
All these calculations are done at the same radius.

Figure 12 displays the profiles of $v_r$, $v_\theta$ and $v_\phi$ along the $\theta$ direction. Note that as discussed in Section 3, the simulation data do not strictly follow the self-similar assumptions, especially for $v_r$ and $v_\theta$, so the profiles for the simulation results are also dependant on the radius. Here we present the profiles at three different radii 10$r_g$, 49$r_g$ and 99$r_g$ (all these are on grid points of the simulation), distinguished by different markers. The steady model results are represented by the solid curves, which are not so smooth as presented in JW11, due to the fact that the calculations are carried out with variable values of $f$ (see Section 4.1). The simulation model is based on pseudo-Newtonian gravity while the steady model is Newtonian, so $v_\mathrm{K}$ is slightly different from each other. However, as we focus on the region between 10 and 100 $r_g$, the difference is negligible.
\begin{figure}[htbp]
\plotone{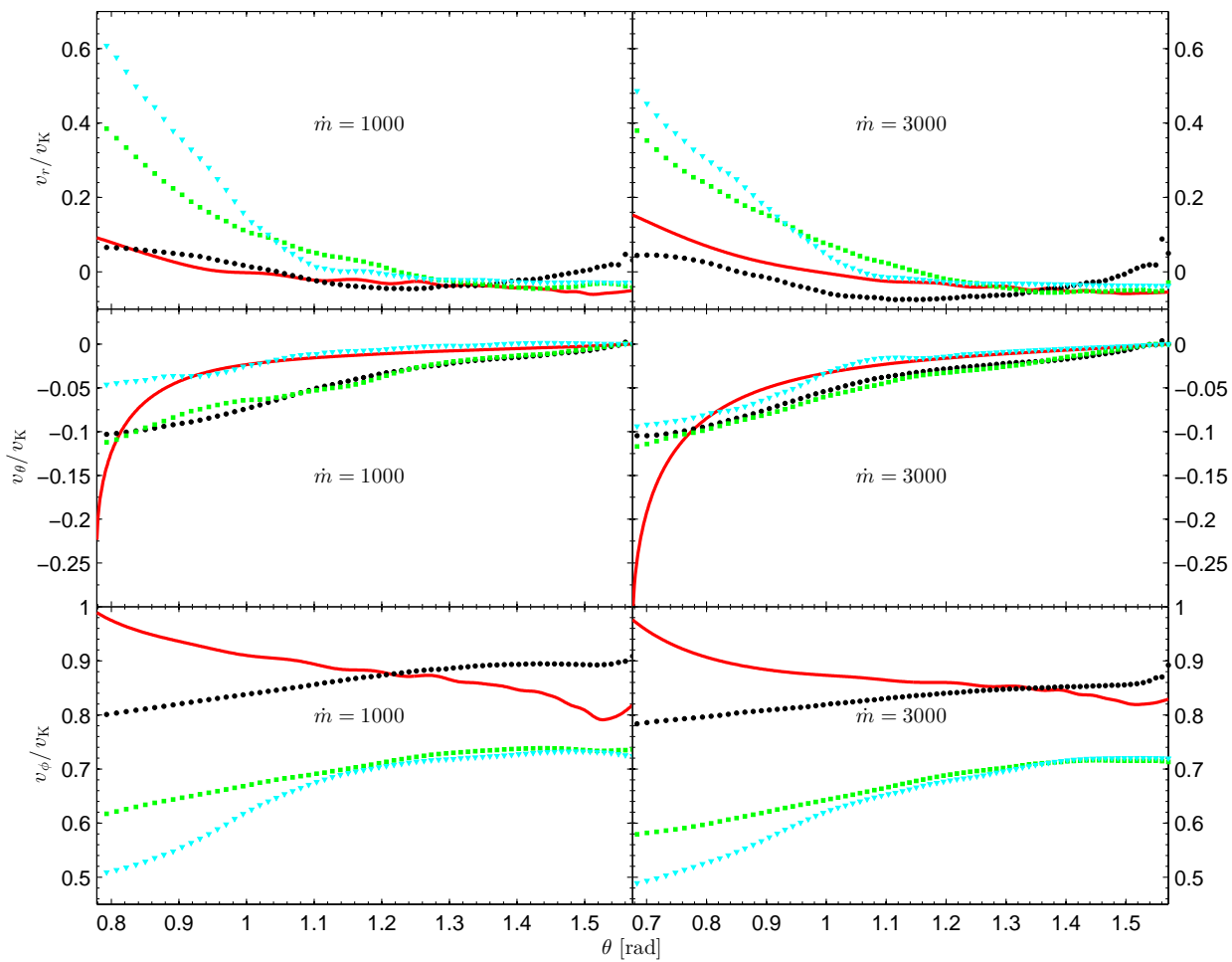}
\caption{
The distribution of $v_r$, $v_\theta$ and $v_\phi$ in $\theta$ direction, all normalized by the local Keplerian velocity. The profiles corresponding to the simulation data are displayed for three different radii of 10$r_g$, 49$r_g$ and 99$r_g$, indicated by black dots, green squares and cyan triangles, respectively (see the online version for the colored figure). The steady model results are represented by solid curves.
}
\end{figure}

As discussed in Section 3, $v_r$ (and partly $v_\theta$) does not follow the self-similar assumptions in the simulation results, likely due to the circular patterns arising from convection. However, the steady model profiles of $v_r$ and $v_\theta$ generally follow the same curve shapes of those obtained from the simulation data, except for the profile of $v_\theta$ near its upper boundary, which arises from the fluctuations of the steady model calculations near the upper boundary. $v_r$ tends to be underestimated in the steady model for large radii, which corresponds to the difference in the flow motion of the velocity field plots mentioned in Section 4.2.

From the fits of $v_\phi$ in Section 3, we know that $v_\phi$ follows the relation $v_\phi \propto r^{-0.65}$ on average for $\dot{m}=1000$ and $v_\phi \propto r^{-0.64}$ for $\dot{m}=3000$. That deviates from the radial scaling of $v_\mathrm{K} \propto r^{-0.5}$, so that although the fitting results of $v_\phi$  {have an average $R^2$ value of 0.99 for both $\dot{m}=1000$ and $\dot{m}=3000$}, the profiles of $v_\phi$ at different radii do not coincide. The profiles of $v_\phi$ in the steady model results tend to increase as the polar angle decreases, which seems to not agree with the simulation results. However, the profile of angular velocity $\Omega$ actually agrees with the simulation results, which will be shown in Section 5.

Figure 13 displays the profiles of $\rho$ and $p$ in the $\theta$ direction. The profiles obtained from the simulation data are shown at three different radii, for the same reason as discussed above. Note that although the radial distribution of total pressure $p$ follows the self-similar assumptions  {well with an average $R^2$ value of 0.99 for both $\dot{m}=1000$ and $\dot{m}=3000$}, as discussed in Section 3, the input parameter $n$ for the steady model is chosen based on the fits of $\rho$, and the exponent is not the same ($n\sim 0.85$ and $0.74$ for $\rho$ while $n\sim 1.03$ and $0.91$ for $p$, for $\dot{m}=1000$ and 3000, respectively). That is why the three profiles of $p$ at different radii of the simulation do not coincide. It can be seen that the profiles from the steady model results generally follow the same curve shape as the simulation profiles. The pressure $p$ near the equatorial plane are overestimated in the steady model results for both accretion rates.
\begin{figure}[htbp]
\plotone{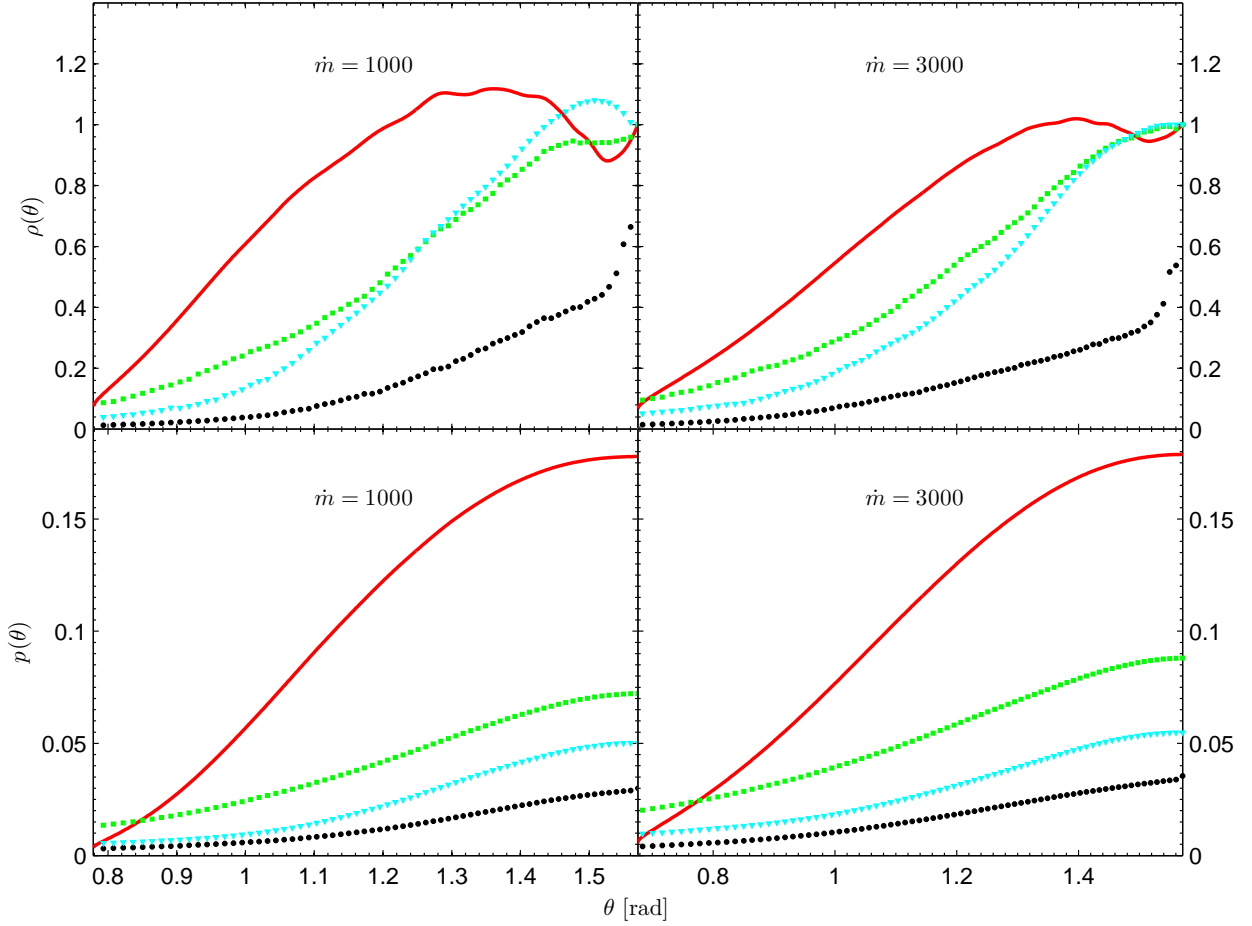}
\caption{
The distribution of $\rho$ (normalized to $\rho(\pi/2)$) and $p$ (normalized by Eq.(23)) in $\theta$ direction. The profiles corresponding to the simulation data are displayed for three different radii of 10$r_g$, 49$r_g$ and 99$r_g$, indicated by black dots, green squares and cyan triangles, respectively (see the online version for the colored figure). The steady model results are represented by solid curves.
}
\end{figure}

Generally speaking, the steady model results agree with the simulation data  {not only qualitatively, but also quantitatively, deviating from the simulation data only several times at most}. Considering that the steady model is based on very simple assumptions and calculations, the result is quite satisfying.
\subsection{Mass, momentum and energy fluxes in outflows}
In this subsection, we investigate the mass, momentum and kinetic energy fluxes in the outflows, calculated from the steady model solutions and the simulation data, as shown in Figure 14. The solid and dashed curves correspond to the fluxes obtained from the simulation data and the steady model solutions, respectively, and the dotted lines correspond to the calculation boundaries in the steady model solutions. In the steady model, the density on the equatorial plane is set to be 1, which can be regarded as normalized by a scale factor. To translate the fluxes obtained from the steady model into real physical units, we need to calculate this scale factor, which is obtained by setting the mass inflow rate at 500 $r_g$ (which is the outer boundary in the simulations) to be the same as the mass supply rate parameter in the corresponding simulation calculations. The momentum flux presented here considers the total momentum, including the $\phi$ component.
\begin{figure}[htbp]
\plotone{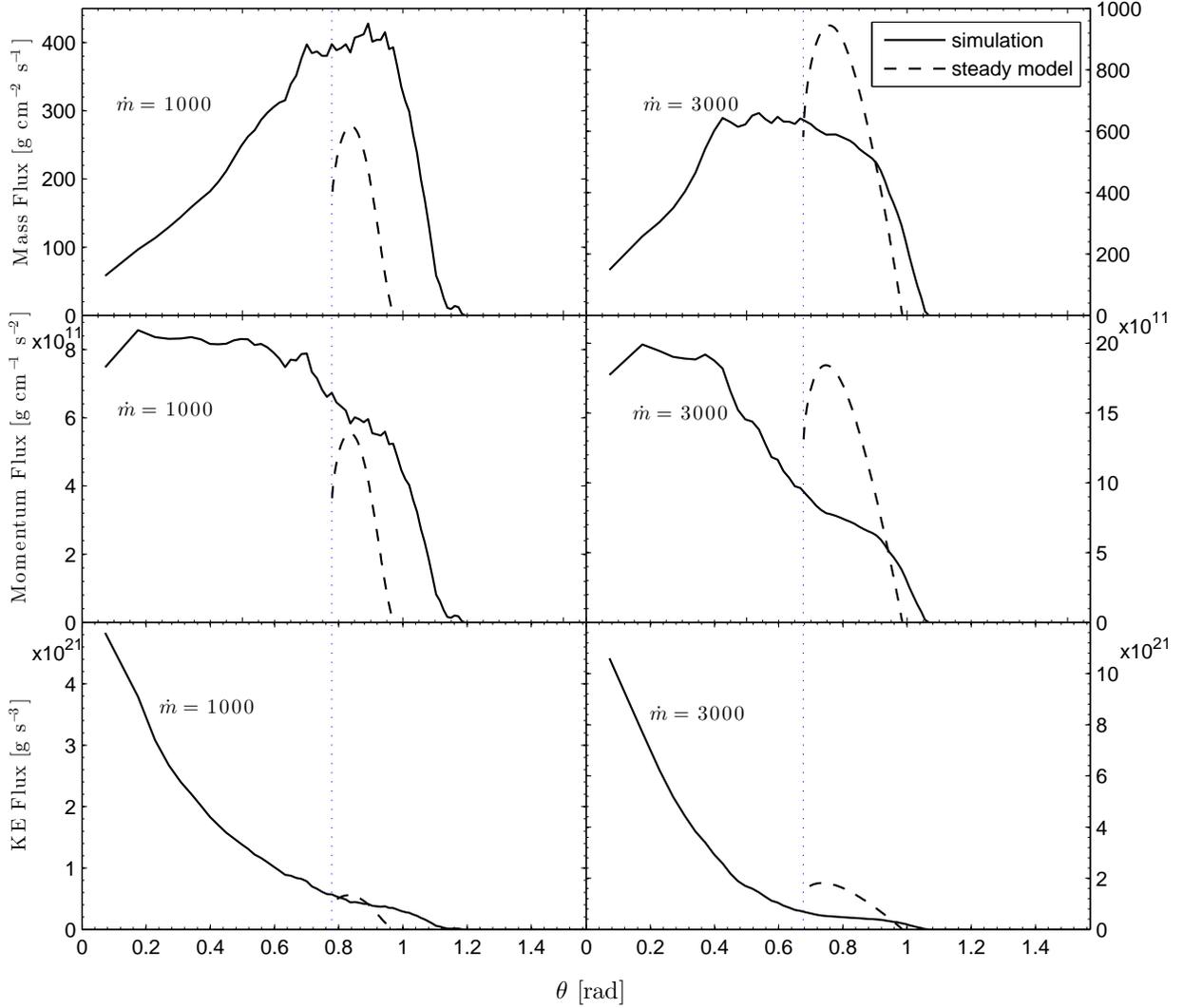}
\caption{Mass, momentum and kinetic energy fluxes in the outflows. The solid and dashed curves correspond to the fluxes obtained from the simulation data and the steady model solutions, respectively. Dotted lines indicate the calculation boundaries in the steady model solutions.
}
\end{figure}

The mass flux obtained from the simulation data peaks at $\theta=0.8914$ rad ($\sim 51.1^\circ$) for $\dot{m}=1000$, and $\theta=0.5381$ rad ($\sim 30.8^\circ$) for $\dot{m}=3000$, respectively. Mass flux also remains high around respective peaks, which means that the outflowing mass is overall directed toward the polar angle around the peak. At the outer boundary of the converged region ($r=100r_g$), $\sim$94\% of the mass inflow are driven away as outflow. The mass flux obtained from the steady model peaks at $\theta=0.8382$ rad ($\sim 48.0^\circ$) for $\dot{m}=1000$ and $\theta=0.7852$ rad ($\sim 43.4^\circ$) for $\dot{m}=3000$, respectively. Both peaks fall in the main outflow region as shown in Fig. 13. There are still significant outflows beyond the calculation boundary of the steady model, and the mass outflow in the calculated region of the steady model only takes up around 9\% of the mass inflow.

For the momentum and kinetic energy fluxes, although those obtained from the simulation data basically agree with those obtained from the steady model in the regions that have been calculated, they keep increasing until near the polar axis, reaching maximum values on the 2nd calculation grid point ($\theta=0.1749$ rad, $\sim 10.0^\circ$) for mass flux, and on the 1st calculation grid point ($\theta=0.0724$ rad, $\sim 4.1^\circ$) for kinetic energy flux. Significant momentum and kinetic energy fluxes exist in the regions around the polar axis that are not calculated by the steady model, and should be taken into consideration if one applies our steady model to observations.

As the mass supply rate increases from $\dot{m}=1000$ to $\dot{m}=3000$, the fluxes appear to increase more for the steady model solutions than for the simulation results. This is likely because that the simulation results have not achieved converged state from 100$r_g$ to 500$r_g$, so that increasing the mass supply rate at the outer boundary has less impact on the outflows at 100$r_g$ in the simulation results.
\section{Discussion}
\subsection{Validity of self-similarity and steady solutions}
The steady model we adopt here is based  {on the assumption of self-similarity}. Physically,  {the assumption is} based on the idea that in the problem, the only length scale of interest is $r$, and the only frequency is $\Omega_\mathrm{K}$, so that components of velocity, as well as the sound speed, should scale with radius as $r\Omega_\mathrm{K}$ (Narayan \& Yi 1995a). The radial distribution of $\rho$ is not quite clear, so it is described with a variable $n$ in the form of $\rho \propto r^{-n}$. Due to the relation that the isothermal sound speed $c_s=\sqrt{p/\rho}$, we can get the radial distribution of $p$ as $p \propto r^{-n-1}$. In the extreme case that $v_\theta$ is assumed to be 0, the value of $n$ can be calculated to be 1.5 via the continuity equation. In this case, all the stream lines of the accreted material would be straight lines pointing directly at the central accretor, and the accretion flow would be composed of pure inflow. For $n<1.5$, outflows will be generated from the accretion disk. $n>1.5$ would indicate that the disk is supplied with mass in the $\theta$ direction as 'inflow wind', which is unlikely to happen in real cases. Self-similar assumptions have been widely adopted in steady accretion disk models (e.g. Narayan \& Yi 1994, 1995a; BB99; BB04; Xue \& Wang 2005; Gu et al. 2009; JW11; Begelman 2012; Gu 2012, 2015; etc.). However, the validity of these assumptions are not quite clear.

In Section 3, we have checked these assumptions with two samples of RHD simulations of supercritical accretion flows. The radial distribution of $v_\phi$ is close to the self-similar form, especially near the equatorial plane, with the exponent of $r$ around -0.6 (Figure 3). The radial distribution of  {$v_r$ and $v_\theta$, however, do not follow the self-similar forms so well. This is mostly} due to the circular patterns in the velocity fields of the simulations. The self-similar forms require that $v_r$ and $v_\phi$ remain in the same direction for the same polar angle $\theta$, therefore the stream lines can not form enclosed rings in the self-similar region, which is obviously violated by the circular patterns. These patterns are not obvious in the time-averaged velocity field plots, but can be seen clearly in a snapshot figure (such as Figure 4 in OMNM05).  {The radial distribution of $v_r$ in the outflows is also influenced by the strong gravitational force near the central black hole, which causes the value of $v_r$ in the outflows to drop significantly near 10$r_g$.} Density $\rho$ and pressure $p$ are in agreement with the relations $\rho \propto r^{-n}$ and $p \propto r^{-1-n}$ as we assume. In our steady model, the mass inflow rate $\dot{M}_\mathrm{inflow} \propto r^{1.5-n}$. The parameter $n$ obtained from the simulation data is smaller for $\dot{m}=3000$ than $\dot{m}=1000$, which implies that stronger outflows are generated when the mass supply rate is increased.

The distribution of velocities, density and pressure in $\theta$ direction calculated by our steady model is basically in agreement with the simulation result, displaying only serval times of difference quantitatively at maximum (see Section 4.3). The largest deviations appear for the $\theta$ profiles of azimuthal velocity $v_\phi$ and pressure $p$. For $v_\phi$, the curve shape seems different from the simulation results in Figure 12. Actually it is not that different, if we show the $\theta$ profiles of angular frequency $\Omega$ instead (Figure 15). In Figure 15, $\Omega/\Omega_\mathrm{K}$ decreases as $\theta$ decreases for both the steady model solutions and the simulation results, which is natural as 'isorotes' (surfaces of constant $\Omega$) tend to be less curved than spheres. Note that although both $v_\mathrm{K}$ and $\Omega_\mathrm{K}$ are set as constants for the same spherical radius, the rotational motion is actually relative to the cylindrical radius, so that the curve shape of $\Omega/\Omega_\mathrm{K}$ is different from that of $v_\phi/v_\mathrm{K}$. Pressure $p$ is overestimated at low latitude and underestimated at high latitude in our model, compared with the simulation data. We conjecture that this arises from the fact that convective motions in the simulation act as an additional mechanism of transporting energy outward, reducing the energy density and consequently the pressure at low latitude and increasing them at high latitude. The convective motion may also play some role in transporting angular momentum, but this is not as clear as the pressure distribution.
\begin{figure}[htbp]
\plotone{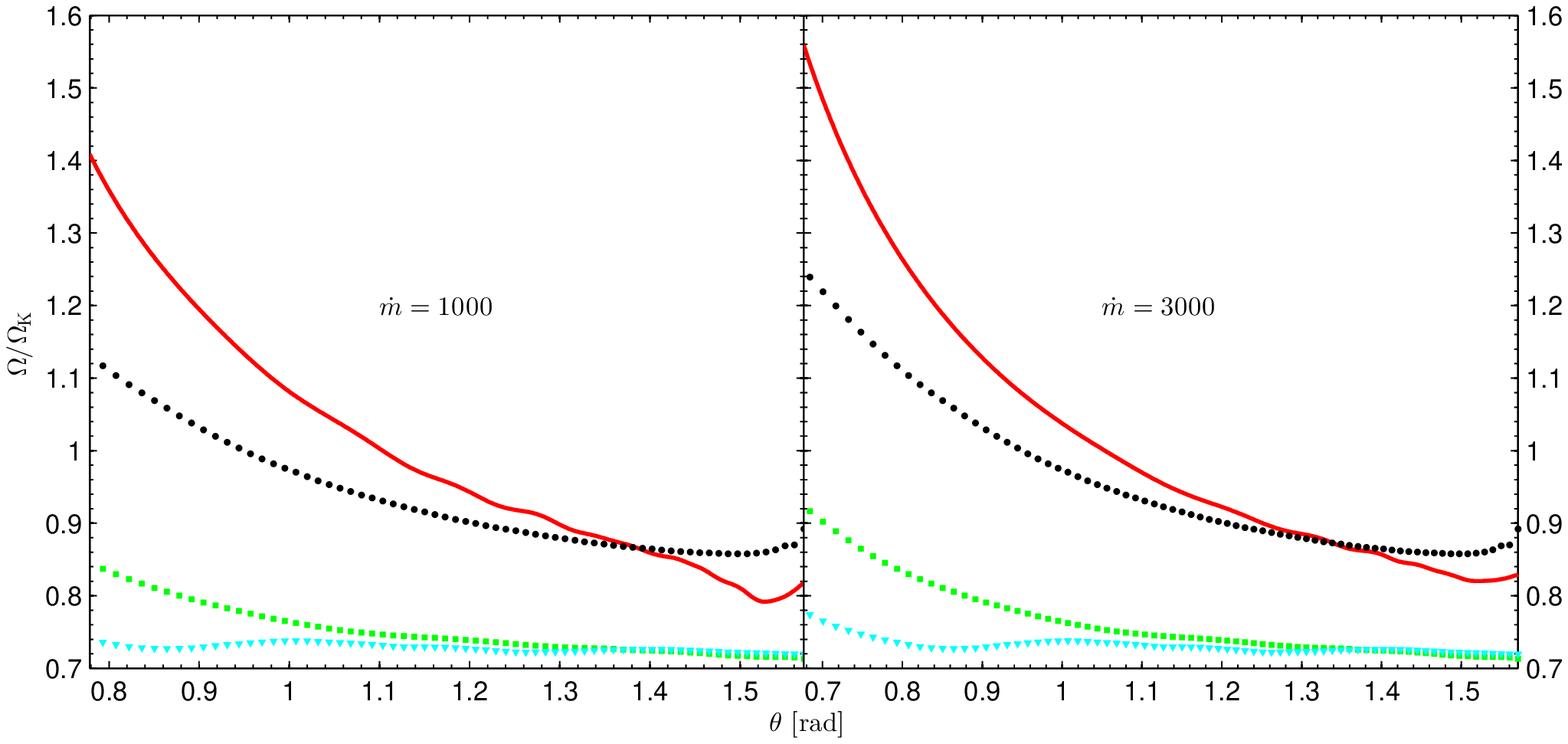}
\caption{
The distribution of $\Omega$ (normalized by corresponding local $\Omega_\mathrm{K}$) in $\theta$ direction. The profiles corresponding to the simulation data are displayed for three different radii of 10$r_g$, 49$r_g$ and 99$r_g$, indicated by black dots, green squares and cyan triangles, respectively (see the online version for the colored figure). The steady model results are represented by solid curves.
}
\end{figure}

It should be noted that the analyses in this paper are based on two samples of RHD simulation runs, and need further confirmation from more simulation data.  {2D radiation-magnetohydrodynamic (RMHD) simulations (Ohsuga et al. 2009; Ohsuga \& Mineshige 2011) and 2.5D special relativistic RMHD simulations (Takahashi \& Ohsuga 2015) of supercritical accretion flows have been performed in recent years, but these simulations are much more time-consuming and thus only have limited calculation time. The general inflow-outflow structure in these simulations is similar to that of our steady model, with inflow near the equatorial plane and outflow above the inflow region. They also show new features such as magnetically collimated jets around the polar axis, which are not described by the steady model. However, limited calculation time, restricted computational domain and the influence of the initial conditions make the current simulations not suitable for detailed comparison, and we leave this topic for future work.}
\subsection{Convection}
In our steady model, it is assumed that the energy transport mechanism is mainly based on advection and radiation, while heat conduction and convection are neglected. While our model is self-consistent, there are certainly other possible solutions with alternative prescriptions. As discussed in the above section, convection seems to have some impact on the structure of supercritical accretion flows. The convective transport is essentially non-local, which acts in large circulation patterns in the snapshots of the simulation data. However, the time-averaged data do not display obvious circulation patterns (see Figure 11). Even so, the effects of momentum and energy transport through convective motions still exist, and need to be considered to improve the agreement between steady solutions and the simulation results, which to some extent represent real astrophysical accretion flows.

In literature, there is a  {series of self-similar models for accretion flows} with convection: the 'adiabatic inflow-outflow solutions' (ADIOS; BB99, BB04; revised in Begelman 2012, which {no longer requires convection to be the dominant mechanism of energy transport}). The BB99 version of ADIOS is a one-dimensional (1D), height-integrated, radially self-similar model of steady radiatively inefficient accretion flows. It is basically a variation of the 1D self-similar ADAF model proposed by Narayan \& Yi (1994), with a variable accretion rate $\dot{m} \propto r^{p}$ ($p$ correspond to $1.5-n$ in our model). This model is expanded and redefined in BB04. They propose that the model is applicable to both 'overfed' (i.e. supercritical) and 'underfed' scenarios of accretion disks and the model is expanded to 2D. The 2D ADIOS model also contains outflow, and poloidal flow in the model is quadrupolar, inward at low latitude and outward at high latitude, which is similar to our model. In their model, convection is included as the main mechanism of energy transport. However, this is not reflected in the energy equation, which is described with polytropic relation (they only require $p \propto \overline{\rho}^\gamma$ when an element of gas changes density) and the energy conservation equation is used to solve for the convective energy flux after the structure of the flow has been obtained. The main postulation about convection in BB04 is that the convective motion is performed along the gyrentropes (surfaces of constant specific angular momentum, Bernoulli function and entropy, which coincide with each other when the disc is marginally stable to the second H{\o}iland criterion), which does not agree with the simulation results here, as convective motions in the simulation data actually operate in large circular patterns.

While both our model and the ADIOS model are based on self-consistent assumptions, our model is more suitable to be applied to the analysis of the simulation data from OMNM05 mainly for two reasons. The first is that, in ADIOS, the energy transport due to radiative flux is assumed to be zero, which is not applicable to the simulations of supercritical accretion flows here (cf. Section 4.1). The transport of energy is mainly due to advection and radiation, which agrees with the assumptions of our steady model. More importantly, in both our model and the simulation data, viscosity still works in the outflow region, while in the ADIOS model it is assumed to no longer work in the outflow part. The outflow-disk model described in BB04 calculates the disk part and outflow part separately. The outflow is assumed to launch from a 'thermal front' where the convective motions quickly dissipate, increasing the entropy of the gas. It is also assumed that in the outflow part, the viscous torque no longer act and the viscous transport of angular momentum and energy also stops. This is quite different from our model, as we calculate the structure of the whole accretion flow together, and the outflow is identified after the structure has been obtained. Obviously our model agrees more with the simulations in this sense. We are also planning to include convection in our steady model in future.

The ADIOS model is revised by Begelman (2012), in which the outflow calculation is much improved, as the outflow structure is no longer assumed to be laminar and inviscid. However, the vertical structure of accretion flow is not obtained in the revised ADIOS model, so it is difficult to compare with our results here. It is suggested in the revised ADIOS model that the mass flux index for supercritical accretion should be less than 1, which corresponds to $n>0.5$ in our model (note that we use different parameter notifications). This agrees with our results in this paper.
\subsection{The 'underfed' case}
As mentioned in introduction, outflows are likely to be generated in both the 'overfed' and the 'underfed' case. We have discussed the 'overfed' case with our steady model and simulation data. However, our steady model is established as a general model which can describe both the 'overfed' and the 'underfed' case by adjusting the input parameters. It is also an interesting topic to compare our steady model with the simulation data in the 'underfed' case, which we are planning to do in future work. Here we just give some general notes in the 'underfed' case.

The 'underfed' case corresponds to a hot, optically thin accretion flow which is radiatively inefficient and thus has an advection-dominated energy transport. Therefore the advective factor can be taken as 1. Radiation pressure is also very small so that $\gamma_\mathrm{equ}$ can be taken as 5/3, which corresponds to gas-pressure dominated accretion (note that some simulations take into account the magnetic pressure, in which case $\gamma_\mathrm{equ}$ should be adjusted accordingly). This leaves $\alpha$ and $n$ to be obtained from simulation data, while the validity of self-similar assumptions should also be checked.

The viscous parameter $\alpha$ can either be obtained from the form of shear stress adopted in hydrodynamical simulations (e.g. Yuan et al. 2012a), or be calculated from simulation data of MHD simulations (e.g. Narayan et al. 2012). The parameter $n$ can be calculated from radial density profile of the simulation data. For example, Yuan et al. (2012a) obtained $\rho \propto r^{-0.65}$, $p \propto r^{-1.7}$, $v_\phi \propto r^{-0.5}$, $v_r \propto r^{-0.55}$ in the case of $\alpha=0.001$, and $\rho \propto r^{-0.85}$, $p \propto r^{-1.85}$ with velocity index around -0.5 (observed from their Figure 4) in the case of $\alpha=0.01$, both for $r\gtrsim 10 r_g$. Narayan et al. (2012) also find $v_r \propto r^{-0.5}$ for $r\gtrsim 10 r_g$. It appears that there are positive evidence for the validity of self-similar assumptions in the 'underfed' case, although detailed investigation still awaits to be made. There are also debates over the importance of convection in the 'underfed' case (e.g. Yuan \& Bu 2010; Narayan et al. 2012; Yuan et al. 2012a, 2012b), which will be investigated in future work.

\section{Summary}
We make comparison between our steady accretion disk model containing outflows and two samples of 2D RHD simulation of supercritical accretion flows. The steady model is based on radial self-similar assumptions of velocity, density and pressure, which are checked with the simulation data. In the converged region of the simulation data, excluding the part too close to the central black hole, azimuthal velocity $v_\phi$, density $\rho$ and total pressure $p$ basically follow the self-similar assumptions, while radial velocity $v_r$ does not, which is likely due to the circular pattern of accretion flow in the $r\theta$ plane of the simulation, caused by convection.  {The radial distribution of $v_r$ in the outflow region is also influenced by the strong gravitational force near the central black hole, which causes the value of $v_r$ in the outflows to drop significantly near 10$r_g$.} Polar velocity $v_\theta$ somewhat follows the self-similar assumptions, although it is also influenced by convection.  {Physically, convection acts as an additional mechanism of transporting momentum other than those considered in our self-similar model, so that profiles of $v_r$ and $v_\theta$ are disturbed and deviate from their self-similar forms. The fact that $v_\theta$ follows self-similarity better than $v_r$ implies that the effects of convection are stronger in the radial direction,} giving us some hint on how to treat convection in future work.

Then we calculate the solutions of the steady model, with input parameters based on the simulation data. In the steady solutions, the distribution of physical quantities in $\theta$ direction basically agree with the simulation results. The agreement is good not only qualitatively, but also quantitatively, as the steady model results deviate only several times from the simulation results at most. The result of comparison is satisfying, considering that the steady model is based on very simple assumptions and calculations. In the simulation results, outflowing mass is overall directed toward polar angle of 0.8382 rad ($\sim 48.0^\circ$) for $\dot{m}=1000$, and 0.7852 rad ($\sim 43.4^\circ$) for $\dot{m}=3000$, and $\sim$94\% of the mass inflow are driven away as outflow (at 100$r_g$), while outward momentum and kinetic energy fluxes are focused around the polar axis. There exist significant mass, momentum and kinetic energy fluxes in the regions that are not calculated by our steady model, and attention should be paid when the model is applied to other theoretical or observational studies. The radial velocity $v_r$ is underestimated at high latitude, and the total pressure $p$ is overestimated at low latitude in the steady model. We conjecture that if convection is included as an additional mechanism of transporting mass, energy and angular momentum, this disagreement may be alleviated. In the two samples of simulation data in this paper, convection is less important than advection and radiation, so our steady model still holds in principle.  Convection may play a larger role in optically thin accretion flows (e.g. Yuan \& Bu 2010, but see Narayan et al. 2012 for a different opinion), which will be investigated in future work. We are also planning to include convection in the steady model in future.

The analyses in this paper are based on two samples of simulation runs, and need further confirmation from more simulation data. As the first step, we compare our steady model with RHD simulation of supercritical accretion flows based on $\alpha p$ prescription of viscosity. As our steady model is parameterized and can be adjusted to correspond to both 'overfed' and 'underfed' accretion flows, we would like to apply our model to simulations of 'underfed' accretion flows in future work. Besides that, there have been advanced MHD simulations recently in which viscosity is generated through magneto-rotational instability (MRI) self-consistently (e.g. Ohsuga et al. 2009; Ohsuga \& Mineshige 2011; Narayan et al. 2012; Yuan et al. 2012b; Takeuchi et al. 2013; etc.). It is not clear whether the results of these complicated simulations can be reproduced by simple analytic models, which will be a topic in our future work.

\acknowledgments

We thank the referee for making very helpful comments and suggestions.

\end{document}